# Online Voltage Stability Assessment for Load Areas Based on the Holomorphic Embedding Method

Chengxi Liu *Member, IEEE*, Bin Wang, *Student Member*, Fengkai Hu, *Student Member, IEEE*, Kai Sun, *Senior Member IEEE*, and Claus Leth Bak, *Senior Member IEEE*

***Abstract*—This paper proposes an online steady-state voltage stability assessment scheme to evaluate the proximity to voltage collapse at each bus of a load area. Using a non-iterative holomorphic embedding method (HEM) with a proposed physical germ solution, an accurate loading limit at each load bus can be calculated based on online state estimation on the entire load area and a measurement-based equivalent for the external system. The HEM employs a power series to calculate an accurate Power-Voltage (P-V) curve at each load bus and accordingly evaluates the voltage stability margin considering load variations in the next period. An adaptive two-stage Pade approximants method is proposed to improve the convergence of the power series for accurate determination of the nose point on the P-V curve with moderate computational burden. The proposed method is illustrated in detail on a 4-bus test system and then demonstrated on a load area of the Northeast Power Coordinating Council (NPCC) 48-geneartor, 140-bus power system.**

## I. Introduction

THE increase of electrical energy demand and the obstacle of concomitant infrastructure upgrading have driven electrical power systems operated closer to their stability limits. At present, electricity utilities are more concerned about the issues of voltage stability. A reliable, accurate, fast, online voltage stability assessment (VSA) is critical in the operational environment for the utilities to not only identify areas vulnerable to voltage instability, especially under stressed system conditions, but also provide system operators with first-hand advices on control actions.

Based on the online state estimation of operating conditions and the dependable power system models, many model-based VSA methods are utilized to estimate the voltage stability margin (VSM) of power systems, such as the modal analysis method [1]-[3], singular value decomposition method [4], [5], voltage sensitivity method [6], [7], bifurcation theory-based method [8], [9] and continuation power flow (CPF) [10]. In CPF, the maximum loadability of a power system can be determined by increasing the system load in a certain way until the maximum loading point is reached. By iteratively performing a prediction-correction scheme, the CPF can preserve the nonlinearity of power flow equations (PFEs) and overcome the singularity problem. However, the computational burden of CPF is still a major concern for online applications, since many iterative computations are required for each scenario [11].

At present, many utilities have also deployed synchronized phasor measurement units (PMUs) on transmission systems to provide the real-time monitoring of voltage stability. Many measurement-based VSA approaches approximate the external system by estimating parameters of a Thevenin equivalent circuit. They rely on real-time synchronized phasor data from PMUs placed on the boundary of a load area [12]-[15]. Because the full observability on all buses in the load area in real time is often unavailable, a widely adopted approach is to offline reduce the load area to either an equivalent load bus or an equivalent simplified network and to online identify the parameters including those for the external system using PMU data. Then, these measurement-based methods for VSA can assess the stability of a load area or the loadability of tie-lines. Therefore, these methods lack the monitoring of each bus in the load area and depend on the accurate identification of equivalent parameters, which is often difficult in the case of fast load variations.

This paper proposes an online hybrid VSA scheme to predict voltage instability of a load area by power flow calculations using a derivative holomorphic embedding method (HEM). An accurate loading limit at each load bus in the load area can be calculated based on online state estimation on the entire load area and a measurement-based equivalent for the external system. A power series in an embedded complex variable on the load level is derived by the HEM. A *physical germ* solution is proposed to ensure that the embedded value of the power series is always equal to the overall loading level of the load area. Therefore, fed by on-line data on active and reactive powers at each load bus, the Power-Voltage (P-V) curve for every bus can be accurately calculated by performing only one-time HEM calculation

This work was supported in part by *NSF CURENT Engineering Research Center* and *NSF grant ECCS-1610025*.

C. Liu, B. Wang and K. Sun are with the Department of EECS, University of Tennessee, Knoxville, TN, USA (email: cliu48@utk.edu, kaisun@utk.edu, bwang@utk.edu).

F. Hu is with Siemens Industry, Inc., Minneapolis, MN, USA (email: fengkai.hu@siemens.com).

C. L. Bak is with the Department of Energy, Aalborg University, Aalborg, Denmark (email: clb@et.aau.dk).

compared to many iterative computations with CPF. Besides, the adoption of adaptive Pade approximants for the power series given by the HEM can guarantee its convergence near voltage collapse, so as to accurately locate the voltage collapse point. Compared with the model-based VSA methods, the proposed scheme has better time performance by using the non-iterative HEM to give the whole P-V curve. Compared with the measurement-based methods, the proposed scheme is able to accurately give stability margin at each bus of the load area thanks to the retained detailed model of the load area.

As a promising non-iterative method for solving PFEs of large power systems, the HEM was first proposed by A. Trias in [16]-[18]. The basic idea of HEM is to design a holomorphic function and adopt its analytic continuation in the complex plane in order to find the power flow solution in a form of power series. The coefficients of power series are calculated in a recursive manner, whose numerical divergence unambiguously signifies the non-existence of solution. The HEM was first demonstrated on systems having only PQ buses and a slack bus [16], and then on systems having PV buses as well [19]-[25]. To enhance the convergence of the derived power series, references [16]-[25] suggest the use of Pade approximants or continued fractions, whose orders are not optimized yet. Reported applications of the HEM include the analysis of saddle-node bifurcation [26], calculating AC/DC systems [27], finding the unstable equilibrium points [28], [29], network reduction [30] and the analysis of limit-induced bifurcation [31].

Although the superiority of the HEM can be seen in literatures, its practical implementation may encounter a precision issue especially for heavily loaded systems [29], [32]. Moreover, these previous HEMs are devoted to find the solution of one specific load condition, so their embedded variables unnecessarily have a physical meaning, e.g. the actual loading level. In [26], the HEM is applied to estimate the saddle-node bifurcation point of static power flow conditions, in which the loads and total generation are scaled at the same rate. Compared with [26], a hybrid VSA approach is applied and demonstrated on a realistic power network, where loads and generation can be scaled independently.

In this paper, the proposed online HEM-based VSA approach can estimate the VSM at each bus of a load area and predict voltage instability. Superior to other power-flow methods, the HEM analytically gives the whole P-V curve connecting the current operating point to the voltage collapse point for an anticipated scenario of load increase such that the VSM is directly obtained. Supplemented with load trending prediction, this approach can predict the trend of voltage for an early warning of voltage instability for system operators. The fast performance of the HEM gives the operators enough time to respond to foreseen instability and take necessary remedial actions.

The rest of this paper is organized as follows. Section II introduces the conventional HEM algorithm. Section III proposes a derivative embedding method starting from a physical germ solution on P-V curves. Section IV introduces the new VSA approach in detail, including the parameter identification of the external grid, online VSA scheme and adaptive Pade approximants method. Section V uses a 4-bus test system to demonstrate the advantages of this new method. Section VI tests the scheme on a load area of the Northeast Power Coordinating Council (NPCC) power system. Finally, conclusions are drawn in Section VII.

## II. Conventional Holomorphic Embedding Load Flow Method

Consider a complex-valued function $x(s)$ of complex variable $s = p+iq$, with real part $p$ and imaginary part $q$. If the embedded complex-valued function $x(p+iq)$ satisfies Cauchy-Riemann equation (1), $x(s)$ is complex-differentiable and thus holomorphic in a neighborhood of the complex $s$-plane [33].

$$i\frac{\partial x}{\partial p}=\frac{\partial x}{\partial q} \tag{1}$$

Under this circumstance, $x(s)$ can be represented in the form of power series (2) in $s$ within its convergence region $\boldsymbol{C}$.

$$x(s)=\sum_{n=0}^{\infty}x[n]s^{n}, s\subset \boldsymbol{C} \tag{2}$$

In order to solve a nonlinear equation of $g(x) = 0$, substituting (2) for $x$ to generate a composite function of embedded variable $s$:

$$g\left(x\right)=g\left[x(s)\right]=0\,. \tag{3}$$

The idea of using the HEM to solve power flows is to embed a complex variable $s$ into the nonlinear PFEs such that in the complex $s$-plane, an analytical solution is originated from a common germ solution and expanded to the objective final solution. Therefore, the power-flow problem becomes how to design an $x(s)$ satisfying the following four criteria:

1) A common *germ* solution at $s = 0$ can be found for the equation $g[x(s)] = 0$. For power flow calculation, the germ solution is conventionally designated as the solution under a no-load, no-generation condition.
2) Ensuring that $g[x(s)] = 0$ also holds at $s = 1$ and the power series (2) can be mathematically induced within a defined number of order $N$, through expanding and equating the coefficients of the same order of $s^n$ in $g[x(s)] = 0$. Thus, the final solution of $x$ is easily obtained by letting $s = 1$ in (2).
3) The $s$-embedded complex function $g[x(s)]$ is required to be analytic continuous (holomorphic) along the path from the germ solution at $s = 0$ to the final solution at $s = 1$.
4) On the path of $s$ before bifurcation occurs, there is no exceptional point (so called branch point) where multiple solutions of $g[x(s)] = 0$ coalesce with each other. For PFEs, these exceptional points only coincide at the bifurcation point.

Consider an $N$-bus system composed of PQ buses, PV buses and slack buses, which are denoted as sets of $\mathcal{P}$, $\mathcal{V}$ and $\mathcal{S}$ respectively. The original PFEs for PQ buses, PV buses and slack buses are expressed in the following (4)-(6) respectively,

$$\sum_{k=1}^{N} Y_{ik}V_k = \frac{(P_i + jQ_i)^*}{V_i^*}, \forall i \in \mathcal{P} \tag{4}$$

$$\begin{cases} P_i = \mathrm{Re}\left(V_i \sum_{k=1}^{N} Y_{ik}^* V_k^*\right) \\ |V|_i = |V_i^{sp}| \end{cases}, \forall i \in \mathcal{V} \tag{5}$$

$$V_i = V_i^{SL}, \forall i \in \mathcal{S} \tag{6}$$

where $P_i$, $Q_i$, $|V_i|$ and $\theta_i$ are the active power injection, reactive power injection, voltage magnitude and phase angle at bus $i$. $V_k$ is the voltage phasor of bus $k$ adjacent to bus $i$. $Y_{ik}$ is the admittance between bus $i$ and bus $k$.

By the HEM formulation, the voltage of each bus and the reactive power of each PV bus are both represented as power series functions of an embedded complex variable $s$, denoted by $V(s)$ and $Q(s)$ respectively. Then, the $s$-embedded equations of PQ buses, PV buses and SL buses in (4)-(6) can be expressed as (7)-(9) respectively. Note that, in order to maintain the holomorphy of $V(s)$, its conjugate $V^*$ should be defined by a separated function as $V^*(s^*)$, not $V^*(s)$. Details of the conventional HEM-based load flow calculation can be found in [21].

$$\sum_{k=1}^{N} Y_{ik,trans}V_k(s) = \frac{sS_i^*}{V_i^*(s^*)} - sY_{i,shunt}V_i(s), \forall i \in \mathcal{P} \tag{7}$$

$$\begin{cases} \sum_{k=1}^{N} Y_{ik,trans}V_k(s) = \frac{sP_i - jQ_i(s)}{V_i^*(s^*)} - sY_{i,shunt}V_i(s) \\ V_i(s)V_i^*(s^*) = 1 + \left(|V_i^{sp}|^2 - 1\right)s \end{cases}, \forall i \in \mathcal{V} \tag{8}$$

$$V_i(s) = 1 + (V_i^{SL} - 1)s, \forall i \in \mathcal{S} \tag{9}$$

## III. New Holomorphic Embedding Method with Physical Germ Solution

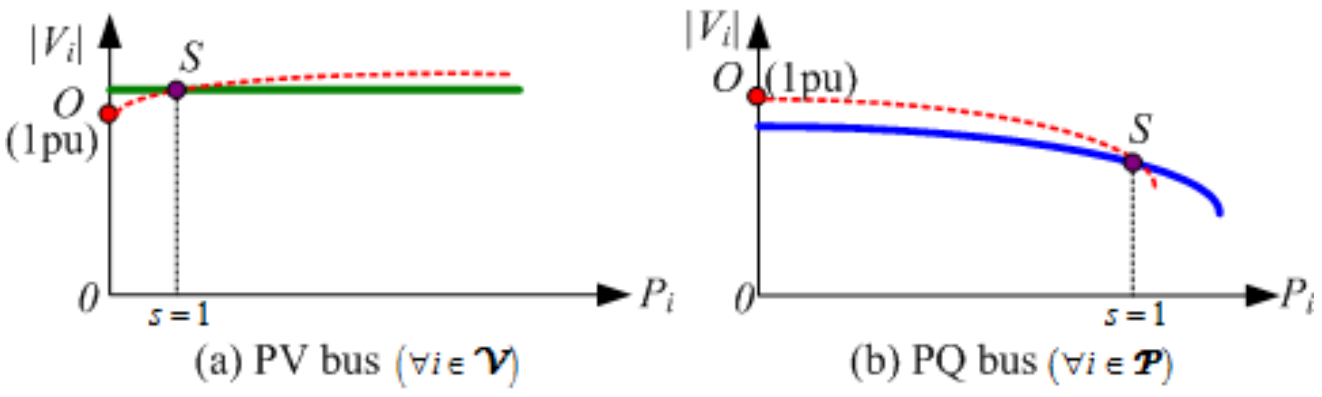

Fig. 1. The procedure of the conventional HEM.

In Eq. (7) and (8), the conventional method decomposes the admittance matrix $Y_{ik}$ into series part $Y_{ik,trans}$ and shunt part $Y_{i,shunt}$ to ensure that the common germ solution has the voltage equal to $1\angle 0^\circ$ under the no-load, no-generation and no-shunt condition at every bus in the network. As illustrated in Fig. 1, HEM intersects the true P-V curve only at $s = 1$, i.e. Point $S$ in Fig. 1(a) and Fig. 1(b) for PV bus and PQ bus respectively. However, the embedded $s$ has no physical meaning at any other values, because $Y_{i,shunt}$ is also scaled by $s$ in (7) and (8).

As illustrated in Fig. 2, different from the germ solution for the conventional HEM, a physical germ solution is proposed in this paper with the no-load no-generation assumption only for PQ buses while non-zero active power is specified and reactive power is injected into PV buses to adjust their voltage magnitudes specifically to $|V_i^{sp}|$.

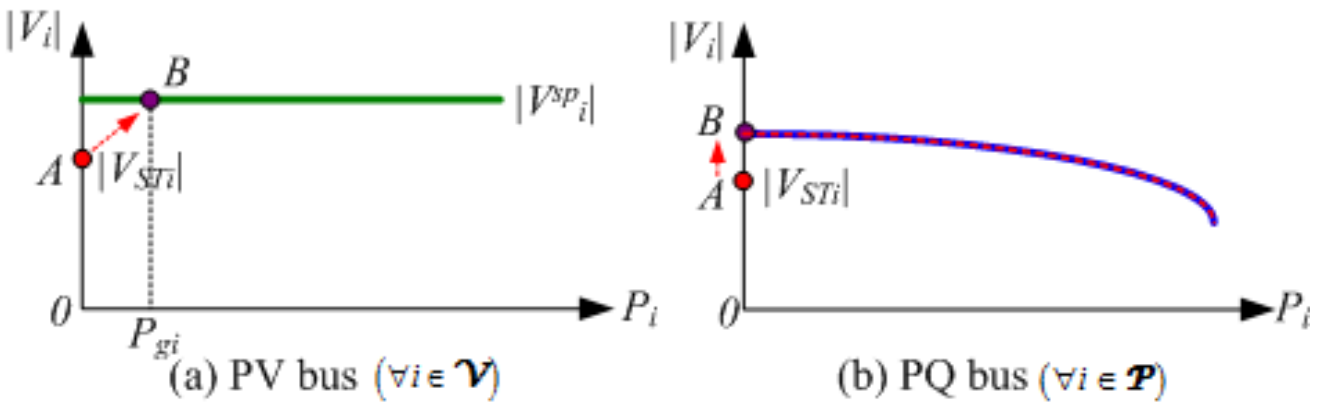

Fig. 2. The procedure of finding the physical germ solution.

Therefore, to find the physical germ, the $s$-embedded equations of PQ buses, PV buses and SL buses in (4)-(6) are expressed as in (10)-(12) respectively, where notations with subscript $gi$ indicates the physical germ solution of bus $i$.

$$\sum_{k=1}^{N} Y_{ik}V_k(s) = 0, \forall i \in \mathcal{P} \tag{10}$$

$$\begin{cases} \sum_{k=1}^{N} Y_{ik}V_k(s) = \frac{sP_{gi} - jQ_{gi}(s)}{V_{gi}^*(s^*)} \\ V_{gi}(s)V_{gi}^*(s^*) = |V_{STi}|^2 + \left(|V_{gi}^{sp}|^2 - |V_{STi}|^2\right)s \end{cases}, \forall i \in \mathcal{V} \tag{11}$$

$$V_{gi}(s) = V_{gi}^{SL}, \forall i \in \mathcal{S} \tag{12}$$

The procedure of finding this physical germ solution consists of two steps. The first step is to find the starting voltage under which the slack bus propagates its voltage to every bus of the network, while all PV and PQ buses have zero injection to the grid, i.e. Point $A$ in Fig. 2(a) and Fig. 2(b) for PV bus and PQ bus, respectively. $V_{STi}$ in (11) represents the starting voltage of bus $i$, which is calculated by (13)-(14).

$$\sum_{k=1}^{N} Y_{ik}V_{STk} = V_i^{SL}, \forall i \in \mathcal{S} \tag{13}$$

$$\sum_{k=1}^{N} Y_{ik}V_{STk} = 0, \forall i \in \mathcal{P} \cup \mathcal{V} \tag{14}$$

Then the second step is to gradually adjust the reactive powers of PV buses to control their voltage magnitudes from the starting voltage $|V_{STi}|$ to the specified voltage $|V^{sp}_i|$ by recursively embedding a series of reactive power equal to $Q_{gi}(s)$. Meanwhile, the active power of each PV bus is fixed at the base value of the original condition $P_{gi}$, i.e. Point $B$ in Fig. 2(a) and Fig. 2(b) for PV bus and PQ bus respectively.

Define the voltage of the physical germ solution for bus $i$ as $V_{gi}(s)$, which is expanded to a power series in $s$.

$$V_{gi}(s) = V_{STi} + V_{gi}[1]s + V_{gi}[2]s^2 + \cdots \tag{15}$$

Then define another power series $W_{gi}(s)$ as the inverse of $V_{gi}(s)$.

$$W_{gi}(s) = \frac{1}{V_{gi}(s)} = W_{gi}[0] + W_{gi}[1]s + W_{gi}[2]s^2 + \cdots \tag{16}$$

Since the starting voltage $V_{STi}$ is calculated by (13)-(14), $W_{gi}[n]$, i.e. the coefficient of term $n$, can be calculated by convolution of the coefficients of terms 1 to $n$-1 of $V_{gi}(s)$ and $W_{gi}(s)$ by

$$W_{gi}[0] = 1/V_{STi} \tag{17}$$

$$W_{gi}[n]V_{STi} + W_{gi}[0]V_{gi}[n] = -\sum_{m=1}^{n-1} W_{gi}[m]V_{gi}[n-m], \quad n \ge 1 \tag{18}$$

Substitute $W_{gi}(s)$ in (16) to the embedded PFEs in (10)-(11)

and expand them as (19) and (20) for PQ buses and PV buses respectively.

$$\sum_{k=1}^{N} Y_{ik}\left(V_{STk} + V_{gk}[1]s + V_{gk}[2]s^2 + \cdots\right) = 0, \forall i \in \mathcal{P} \quad (19)$$

$$\begin{cases} \sum_{k=1}^{N} Y_{ik}\left(V_{STk} + V_{gk}[1]s + V_{gk}[2]s^2 + \cdots\right) \\ = \left[sP_{gi} - j\left(Q_{gi}[1]s + Q_{gi}[2]s^2 + \cdots\right)\right]\left(W_{gi}^*[0] + W_{gi}^*[1]s + \cdots\right) \\ \left(V_{STi} + V_{gi}[1]s + \cdots\right)\left(V_{STi}^* + V_{gi}^*[1]s + \cdots\right) \\ = \left|V_{STi}\right|^2 + \left(\left|V_{gi}^{sp}\right|^2 - \left|V_{STi}\right|^2\right)s \end{cases}, \forall i \in \mathcal{V} \quad (20)$$

Equating the coefficients of $s$, $s^2$,… up to $s^n$ on both sides of (19) and (20), $V_{gi}[n]$ and $Q_{gi}[n]$ are obtained by the terms 0 to $n$-1 of $W_{gi}(s)$ and $Q_{gi}(s)$ from (21) and (22).

$$\sum_{k=1}^{N} Y_{ik} V_{gk}[n] = 0, \forall i \in \mathcal{P} \quad (21)$$

$$\begin{cases} \sum_{k=1}^{N} Y_{ik} V_{gk}[n] = P_i W_{gi}^*[n-1] - jQ_{gi}[n] W_{gi}^*[0] - j\left(\sum_{m=1}^{n-1} Q_{gi}[m] W_{gi}^*[n-m]\right) \\ V_{gi}[n] V_{gi}^*[0] + V_{gi}[n] V_{gi}^*[0] = \varepsilon_i[n-1] \end{cases}, \forall i \in \mathcal{V} \quad (22)$$

$\varepsilon_i[n-1]$ defined by (23) will quickly converge with a small number of terms $n$, since it contains the high order terms of $V_{gi}[n]s^n$.

$$\varepsilon_i[n-1] = \delta_{n1} \cdot \frac{1}{2}\left(\left|V_{gi}^{sp}\right|^2 - \left|V_{STi}\right|^2\right) - \frac{1}{2}\left(\sum_{m=1}^{n-1} V_{gi}[m] V_{gi}^*[n-m]\right) \quad (23)$$

$\delta_{nj}$ in (23) denotes the Kronecker delta function that the coefficient occurs only for order of $n = j$ and vanishes for other orders.

$$\delta_{nj} = \begin{cases} 1 & \text{if n = j} \\ 0 & \text{otherwise} \end{cases} \quad (24)$$

In (22), $V_{gi}[n]$ and $W_{gi}[n]$ are unknown complex numbers, and $Q_{gi}[n]$ is an unknown real number. Move all unknowns of the $n$th order coefficients to the left hand side and break the PFEs into real and imaginary parts. Then, a matrix equation is created containing all $V_{gi}[n]$, $W_{gi}[n]$ and $Q_{gi}[n]$. There are 5 real unknowns in total for each PV bus, as $V_{gi}[n]$ and $W_{gi}[n]$ are complex values and $Q_{gi}[n]$ is a real number. Assume that there are $l$ slack buses, $m$ PQ buses and $p$ PV buses in the $N$-bus network. Then, the dimension of the matrix equation equals to $2l+2m+5p$. The matrix equation to find the physical germ solution of a demonstrative 3-bus system is introduced in detail in APPENDIX-A.

After obtaining the physical germ solution, the final process of the proposed HEM is similar to the conventional HEM. Table I shows the embedding method of HEM with the proposed physical germ solution for PQ, PV and SL buses, respectively, where $s$ represents the loading level only for PQ buses. The difference mainly lies on the right hand side of PQ bus equation in Table I, that $s$ is multiplied by the complex load of the PQ bus. Note that, different from the conventional HEM during the process of embedding, the active powers of PV buses are not multiplied by $s$, indicating invariant generation outputs under load increase. However, if frequency regulation is considered with PV buses, which typically models generator buses, an additional term with generation regulation factor $\alpha_i$ can be attached as shown in (25) to represent the effect of frequency regulation. Here $\alpha_i P_i$ represents the increase rate of active power of generator $i$ with respect to the overall load variation.

$$\sum_{k=1}^{N} Y_{ik} V_k(s) = \frac{\left(P_{gi} - jQ_{gi}\right) + s\left(\alpha_i P_i\right) - jQ_i(s)}{V_i^*(s^*)}, \forall i \in \mathcal{V} \quad (25)$$

TABLE I.
SOLUTIONS OF PFEs FOR DIFFERENT BUS TYPES WITH THE PROPOSED HEM

| Type | Physical germ solution (s = 0) | Final solution (s ≠ 0) |
|---|---|---|
| PQ | $\sum_{k=1}^{N} Y_{ik} V_k(s) = 0$ | $\sum_{k=1}^{N} Y_{ik} V_k(s) = \frac{s(P_i - jQ_i)}{V_i^*(s^*)}$ |
| PV | $\sum_{k=1}^{N} Y_{ik} V_k(s) = \frac{P_{gi} - jQ_{gi}}{V_{gi}^*}$ <br> $V_i(s) V_i^*(s^*) = \left\|V_{gi}\right\|^2$ | $\sum_{k=1}^{N} Y_{ik} V_k(s) = \frac{(P_{gi} - jQ_{gi}) - jQ_i(s)}{V_i^*(s^*)}$ <br> $V_i(s) V_i^*(s^*) = \left\|V_{gi}\right\|^2 + \left(\left\|V_i^{sp}\right\|^2 - \left\|V_{gi}\right\|^2\right)s$ |
| SL | $V_i(s) = V_i^{SL}$ | $V_i(s) = V_i^{SL}$ |

## IV. Online Voltage Stability Assessment

### A. Identification of External System Parameters

For a load center fed by an external system through multiple tie lines, those tie lines may have different power transfer limits in terms of voltage stability, especially in the case that their couplings are weak. Like the $N$+1 buses equivalent system proposed in [15], the external system is regarded as a single voltage source $E$ connected to $N$ boundary buses of the load area respectively by $N$ branches with impedances $z_{E1}$ to $z_{EN}$, as shown in Fig. 3. Thus the coupling relationship among tie lines and boundary buses are retained. Unlike that $N$+1 buses equivalent, all buses in the load area are retained in this paper and their VSMs will be calculated by the proposed HEM.

Sequential quadratic programming (SQP) method is applied to identify the parameters of external grid, i.e. $E$ and $z_{E1}$ to $z_{EN}$, using synchronized data of complex power $S_i$ and voltage phasor $V_i$ measured at the boundary buses [34].

Assume that a time window of $K$ measurement points is obtained by PMUs. $V_i(k) = \left|V_i(k)\right| \angle \theta_i(k)$ and $S_i(k) = P_i(k) + jQ_i(k)$ are defined respectively as the received complex power and voltage phasor at boundary bus $i$ at time point $k$. Therefore, the parameters of the external system can be obtained by solving the following optimization problem,

$$\min J^{ex} = \sum_{k=1}^{K} \sum_{i=1}^{N} \frac{\omega_e}{N}\left[e_i^{ex}(k)\right]^2 + \sum_{i=1}^{N} \omega_z \left(\frac{r_{Ei}}{r_{EP}} - 1\right)^2 + \sum_{i=1}^{N} \omega_z \left(\frac{x_{Ei}}{x_{EP}} - 1\right)^2$$
$$\text{s.t. } E > 1, r_{Ei} \geq 0 \quad (26)$$

where the 1st term gives the estimation errors for power flow equations for all the time instants. The error at time instant $k$ is defined as (27)

$$e_i^{ex}(k) = E - \left|\left(P_i(k) - jQ_i(k)\right)\left(r_{Ei} + jx_{Ei}\right) + \left(V_i(k)\right)^2\right| / V_i(k) \quad (27)$$

The 2nd and 3rd terms summate normalized differences in resistance $r_{Ei}$ and reactance $x_{Ei}$ of $z_{Ei}$ between the estimates for the current and previous time windows. $\omega_e$ and $\omega_z$ are the weighting factors respectively for variances of $E$ and $z_{Ei}$ over the time window.

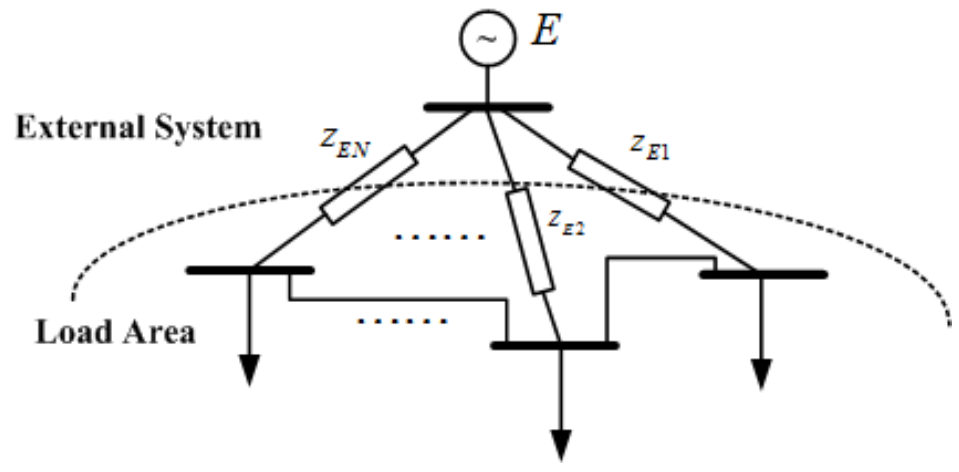

Fig. 3. Reduction of the external system to a voltage source.

As the equivalent parameters of the external system are updated in real time by solving the above optimization problem, the HEM-based online VSA scheme can be performed directly on the reduced system in Fig. 3, while retaining all load buses in the load area. Compared with other equivalent methods, this method retains the couplings among tie lines and boundary buses.

### B. Online Voltage Stability Assessment

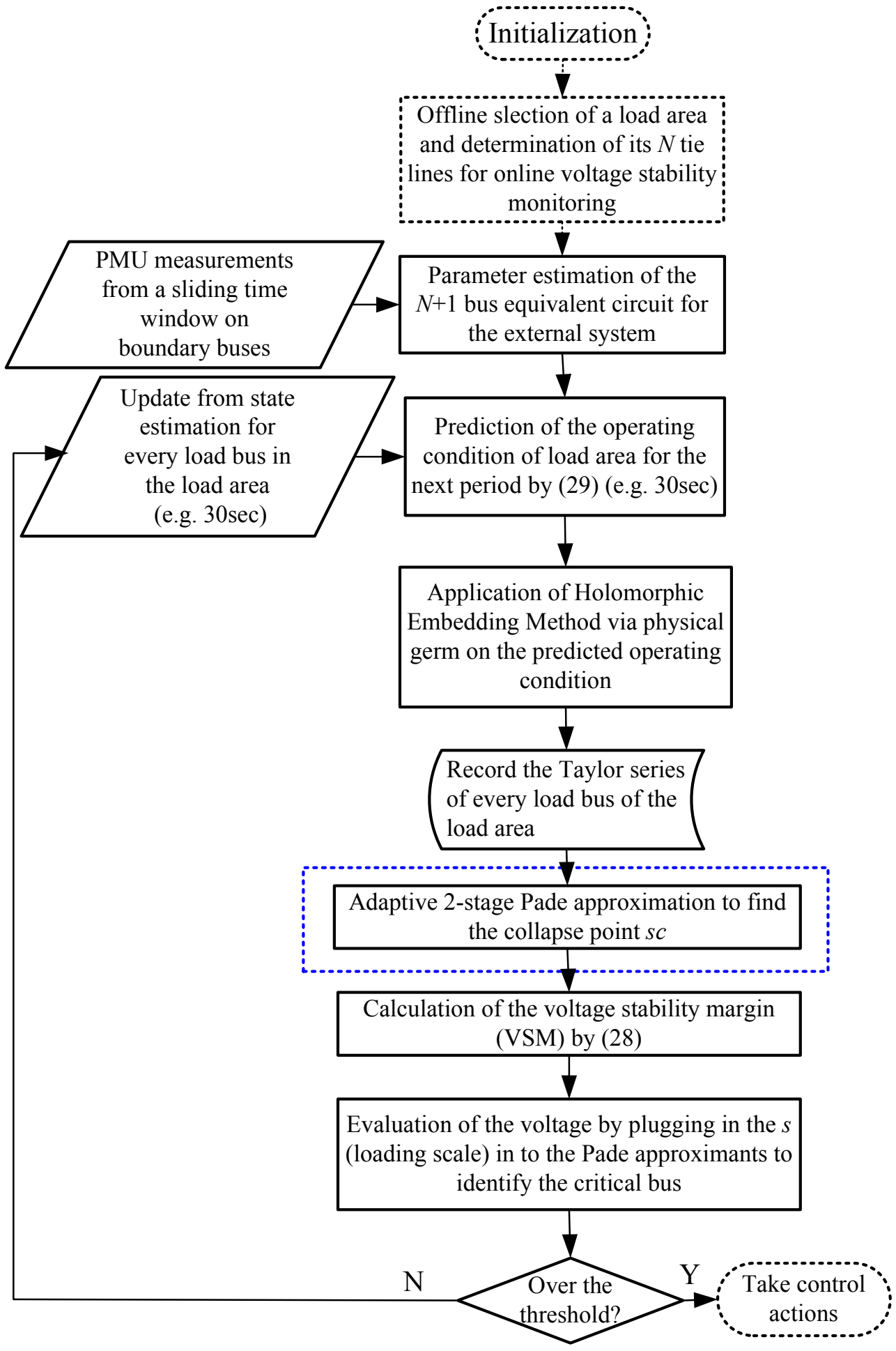

Fig. 4. Flowchart of the HEM-based online VSA scheme.

As shown by the flowchart of the HEM-based online VSA scheme in Fig. 4, it offline selects a load area and identifies all $N$ branches connecting the load area with the rest of the power system. Then, fed by the synchronized PMU measurements for a sliding time window, the parameters of the equivalent circuit can be estimated online to represent the external grid. Meanwhile, the state of each bus in the load area is updated from the state estimation. It can be assumed that the active power, reactive power and voltage of every bus of the load area are obtained from the state estimator, e.g. every 30 seconds [35]. Practically, the loads may vary randomly in a load area, so load trending prediction for the next 30 seconds is adopted. Applying the forecasted loading condition and identified parameters of the equivalent external system, the proposed HEM generates the power series for every bus regarding the overall loading scale of the load area. Further, an adaptive 2-stage Pade approximants method is used to find the accurate collapse point and then the corresponding VSM can be calculated. Additionally, the voltage of each bus with respect to overall load increase can be evaluated by plugging in the loading scale into the analytical expression of the Pade approximants. Hence, the critical bus with the lowest voltage can be identified. Finally, if the voltage margin for the next period violates a pre-specified threshold, control actions may be taken immediately on that critical bus.

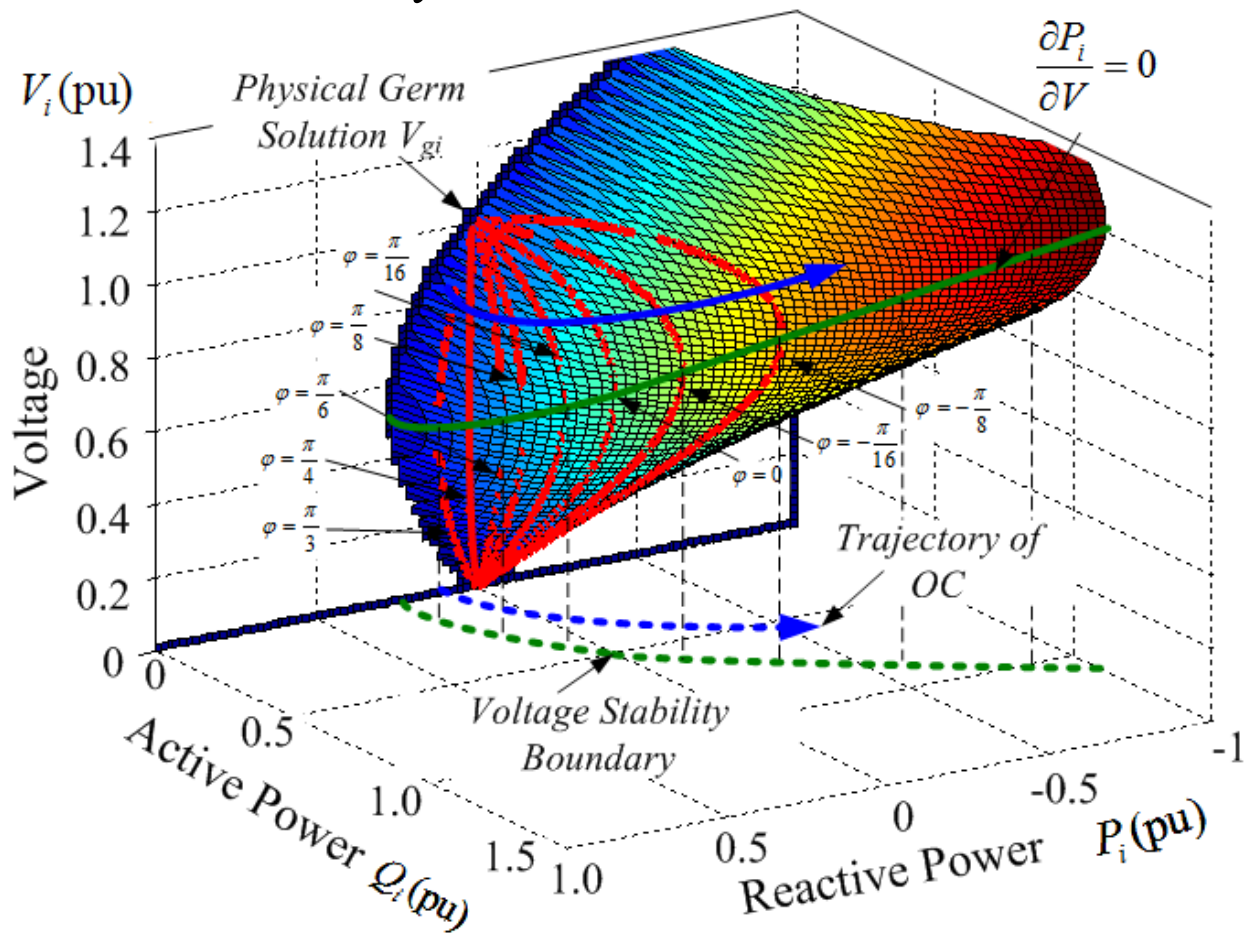

Fig. 5. Illustration of the voltage w.r.t. its active and reactive powers and the voltage stability boundary for a load bus.

Fig. 5 illustrates how the voltage magnitude of a load bus $i$ changes with its active and reactive powers. The physical germ solution $V_{gi}$ clearly shows the no-load condition of the bus. The green line highlights the voltage stability boundary of the load bus with respect to different power factors, while the arrowed blue line illustrates the trajectory of the operating condition (OC) at bus $i$.

Fig. 6 depicts the projection of Fig. 5 onto the P-Q plane to illustrate the procedure of this HEM-based online VSA scheme. The $x$-axis and $y$-axis represent the active and reactive powers of bus $i$. The voltage stability boundary represents the voltage collapse point with respect to different power factor angles $\varphi$. The point $OC_0$ is the original operating condition with $s_0$ = 1. After implementation of this proposed HEM, the P-V curve can be obtained by plugging various values of $s$ into the power series of $V(s)$ under the assumption that all loads are scaled up simultaneously at the same rate and the power factor of each load is maintained invariant.

Then adaptive Pade approximants help to accurately predict the voltage collapse point $sc_t$, which will be introduced in the next sub-section. The VSM of the overall system is defined in (28), where $sc_t$ represents the maximum loading scale at time $t$ until the voltage collapses. VSM physically means the maximum limit of the loading scale in percentage to maintain the voltage stability of the load area.

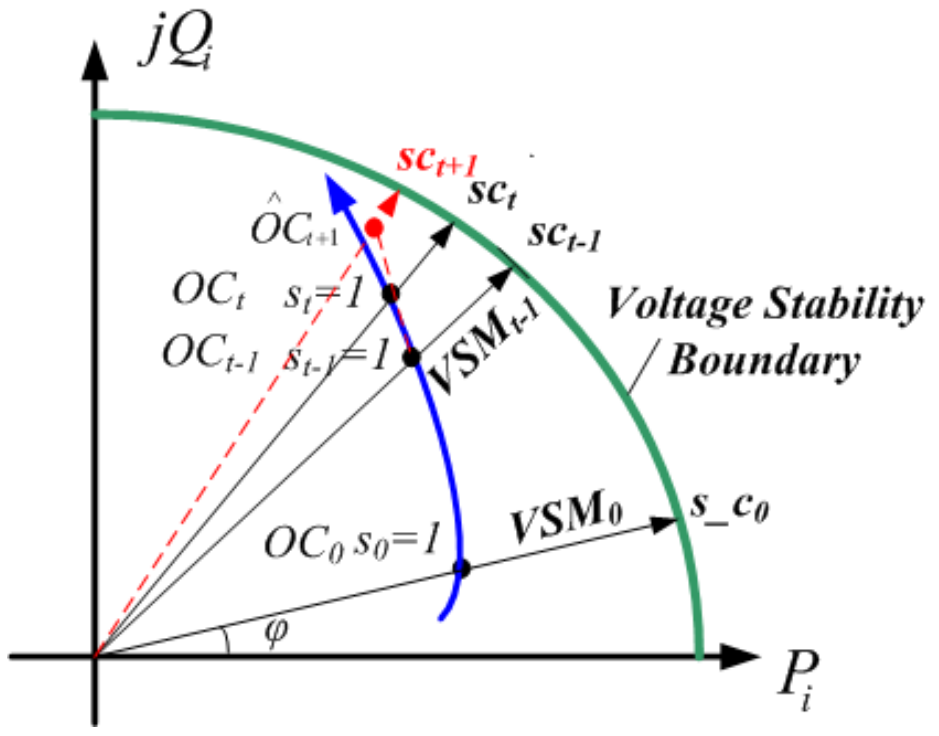

Fig. 6. Illustration of HEM-based VSA scheme.

As shown in Fig. 6, assuming the current operating condition (at time $t$) is the point $OC_t$, the load trending prediction for the next 30 seconds period is adopted, which is based on the linear extrapolation of previous periods. In (29), $S_{i,t}$ denotes the complex power of bus $i$ at time $t$.

$$VSM_t = (sc_t - 1) \times 100\% \tag{28}$$

$$S_{i,t+1} = S_{i,t} + \Delta S_{i,t} = 2S_{i,t} - S_{i,t-1} \tag{29}$$

### *C. Adaptive Two-stage Pade approximants*

As mentioned in the previous section, voltage can be expressed in the form of power series (30) by the HEM. However, the precision issue could have a non-ignorable impact on the performance of HEM, especially when the loading level $s$ extends to the heavily stressed conditions [29], [32]. The reason of this problem lies in the limited convergence region of a truncated power series with limited arithmetic precision. In addition, the double-precision calculation with about 16-digits practically becomes exhausted to decrease the errors of PFEs to 1$e$-13 or lower. An alternative solution is to increase the arithmetic precision to much more digits, i.e. 2000 digits in [32], but the convergence region still cannot reach the actual stability boundary.

From Stahl's Pade convergence theory in [36] and [37], the diagonal or close to diagonal Pade approximants converge to the original function in the maximum domain if the original function is holomorphic. In other words, Pade approximants have the best convergence performance with equal or nearly equal orders between the numerator and denominator, i.e. $|L-M| \leq 1$ in (31). Different from the conventional HEM using Pade approximants to determine the convergence of PFEs [16]-[18], an adaptive two-stage algorithm is proposed here to find the optimal order of Pade approximants for each voltage. Viskovatov's method is adopted here to find the coefficients of Pade approximants [38]. This is carried out with the double precision arithmetic computation after the HEM is performed.

$$V_i(s) = \sum_{n=0}^{N} V_i[n] s^n \tag{30}$$

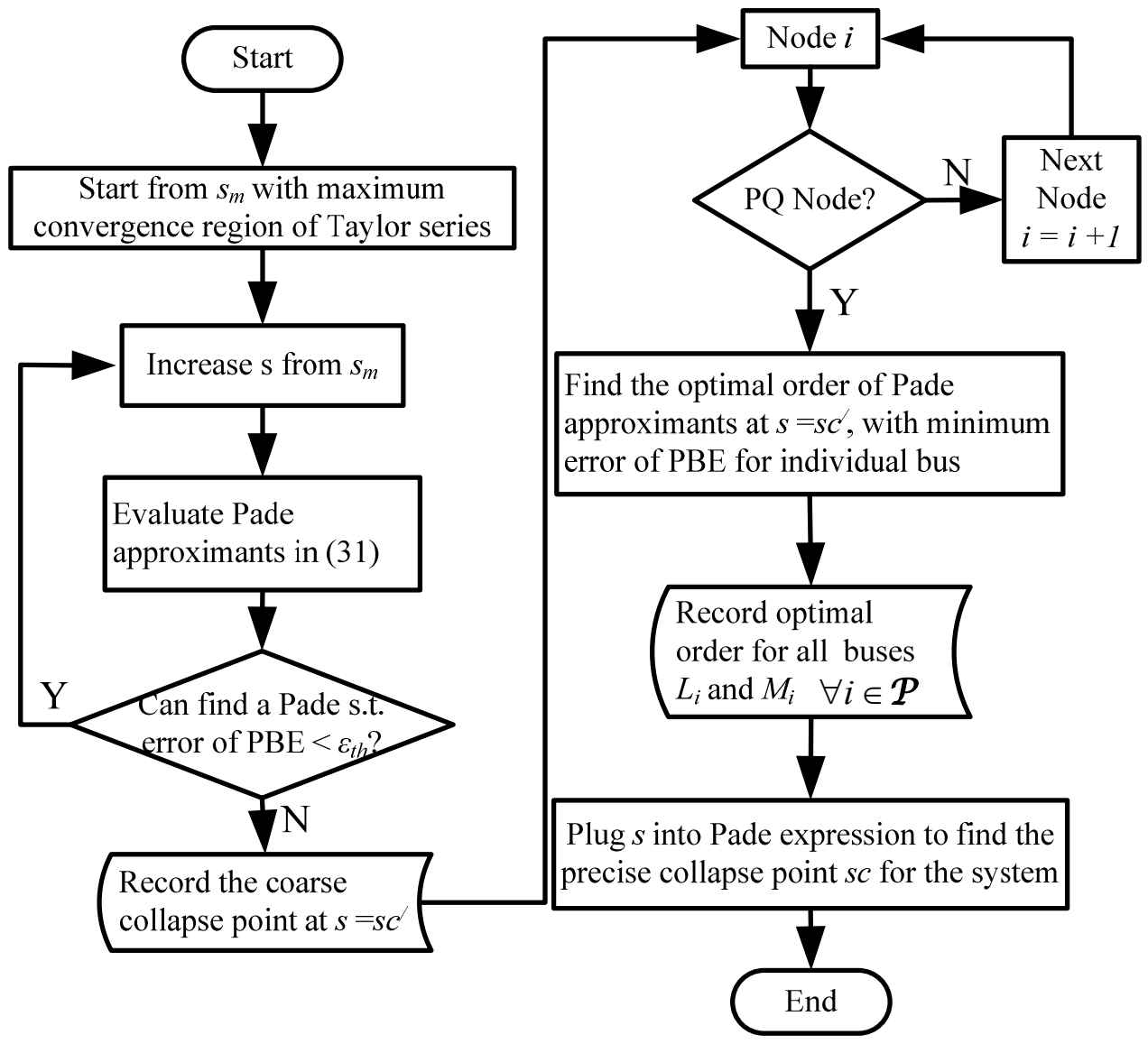

Fig. 7. Flowchart of adaptive two-stage Pade approximants to find the collapse point of the power system.

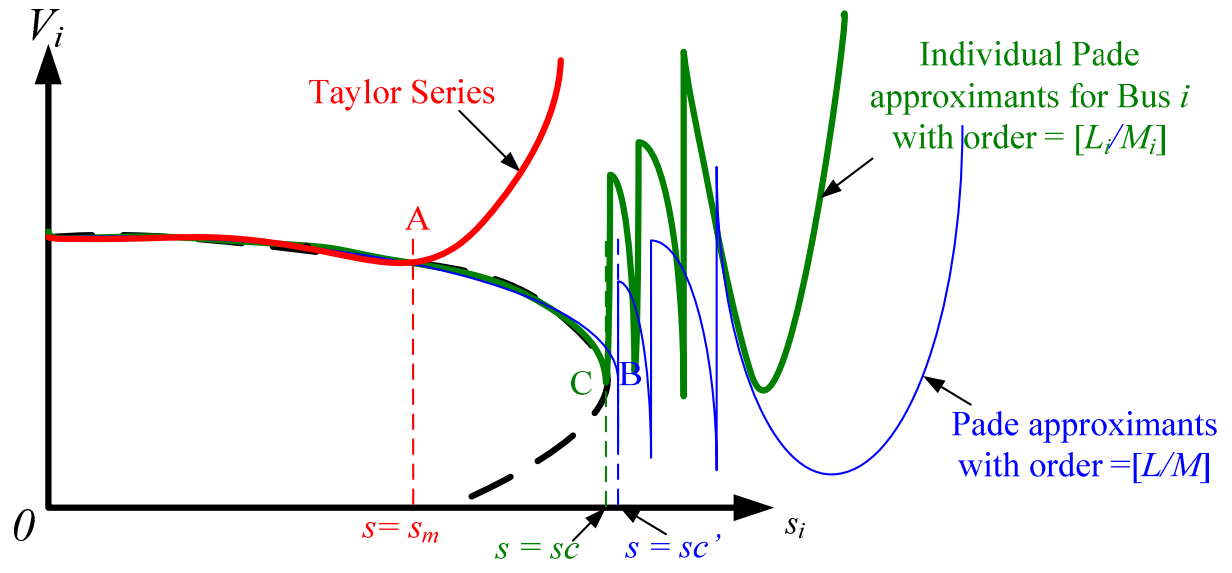

Fig. 8. Illustration of adaptive two-stage Pade approximants.

$$V_i(s) = \sum_{l=0}^{L} a_l s^l \Big/ \sum_{m=0}^{M} b_m s^m = \frac{a_0 + a_1 s + a_2 s^2 + \cdots + a_L s^L}{b_0 + b_1 s + b_2 s^2 + \cdots + a_M s^M} \tag{31}$$

As shown by the flowchart in Fig. 7, the adaptive Pade approximants method consists of two stages, i.e. finding a coarse collapse point at $sc'$ and then identifying a precise collapse point at $sc$. The first stage is to plug values of $s$ into the power series about PQ buses created by the HEM to find the maximum convergence region at $s = s_m$, shown as point A in Fig. 8. Then the full order Pade approximants, i.e. $[L/M]$ for the power series can improve the convergence region from $s = s_m$ to the coarse collapse point $s = sc'$, where the maximum $\varepsilon(s)$ of all PQ buses can be less than a preset threshold $\varepsilon_{th}$, shown as point B in Fig. 8. This process is defined by equation in (32).

$$sc' = \max s, \text{ s.t.}$$
$$\varepsilon(s) = \max_i \left| \sum_{k=1}^{N} Y_{ik} V_k(s) - \frac{s S_i^*}{V_i^*(s^*)} \right|_{i \neq k} < \varepsilon_{th}, \forall i \in \mathcal{P} \tag{32}$$

The second stage is to find the optimal order of Pade approximants for individual bus $i$ to achieve the minimum error of PBEs at $s = sc'$ (31). The optimal order for individual bus $i$ is recorded as $L_i$ and $M_i$, so the Pade expression of voltage is obtained by truncating $[L_i/M_i]$ orders in (31) while still satisfying $|L_i-M_i| \leq 1$. The collapse point of bus $i$ is the nearest pole of the truncated Pade approximants, i.e. $sc_i$.

Finally, the final collapse point predicted at time $t$, i.e. $sc_t$ in (28), is found by selecting the minimum $sc_i$ of all buses, shown by point C in Fig. 8.

$$sc_t = \min_i sc_i = \min_i \left( \min_{M_i \& L_i} \varepsilon(sc') \right), \forall i \in \mathcal{P} \qquad (33)$$

This adaptive two-stage method finds a good tradeoff between the error tolerance in the solution and the condition number of the Pade matrix. When evaluating a diagonal $[L/M]$ Pade, coefficients of terms up to $s^{L+M}$ are required for the power series. The condition number of the Pade matrix increases as the degree of the diagonal Pade approximants increases. Due to the adopted double-precision arithmetic and the round-off errors in the calculation process, evaluating the highest degree of Pade approximants usually means adding inaccurate numbers very close to zero to the numerators and denominators of (31), which leads to an inaccurate estimation of collapse point [21]. Therefore, adaptively adjusting to an appropriate degree of Pade approximants for each load bus can increase the accuracy.

### *D. Considering Different Load Models*

Even for steady-state analysis, load models are important in the VSA especially when voltage dependent loads exist. For example, the actual voltage collapse point no longer coincides with the nose point of the P-V curve when load cannot be modelled as 100% constant power load. A new power series about the voltage magnitude, i.e. $M_i(s) = |V_i(s)|$, should be integrated to represent the dependence between the loading level and the voltage magnitude. When ZIP load is considered, the actual load varies nonlinearly with loading level $s$. APPENDIX-B demonstrates the embedding method and the final matrix equation on a 3-bus system with ZIP load. For dynamic loads such as induction motors, differential equations would have to be included besides the PFEs, which is not the focus of this paper.

### *E. Computational burden of the Proposed Scheme*

After the data collected from the state estimation, the ideal computation time of the proposed VSA scheme includes three parts, i.e. (i) the time $T_{Germ}$ for finding the physical germ solution, (ii) the time $T_{PS}$ for finding the power series by the HEM using the physical germ solution, and (iii) the time $T_{Pade}$ for calculating the collapse point by the adaptive Pade approximants.

$$T_{HEM} = T_{Germ} + T_{PS} + T_{Pade} \qquad (34)$$

$$T_{Germ} = N_{Germ} \cdot t_M \qquad (35)$$

$$T_{PS} = N_{PS} \cdot t_M \qquad (36)$$

Since each term is calculated by a matrix multiplication and several convolutions, the computation times $T_{Germ}$ and $T_{PS}$ mainly depend on the numbers of terms required to meet the error tolerance, i.e. $N_{Germ}$ and $N_{PS}$ in (35) and (36), respectively. $t_M$ indicates the average computation time for each term. Less error tolerance implies more terms in the power series to approximate the P-V curve. Practically, the maximum number of terms is in the range of 40-60 since the float point precision can be exhausted [16]. The computation burden of $T_{Pade}$ mainly depends on the number of PQ buses in the network.

Compared with the CPF, which starts from a certain power flow solution and linearly predicts and corrects an adjacent point on the P-V curve step by step using the Newton-Raphson method, the proposed HEM has better time performance. Although it also starts from a certain point, the HEM predicts the P-V cure nonlinearly and directly to the branch point [11]. The computation time of CPF and the proposed method will be compared in the following sections.

## V. Demonstration on a 4-Bus Power System

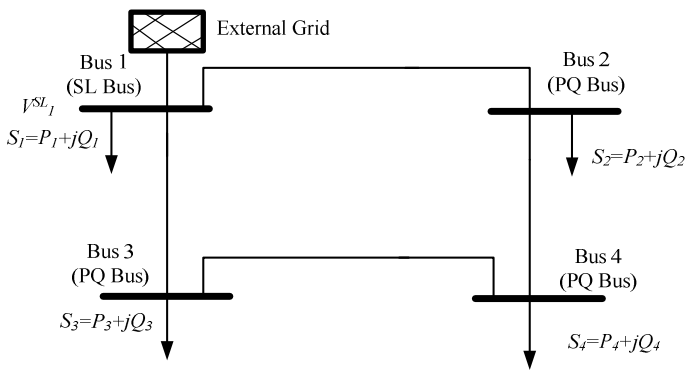

Fig. 9. One-line diagram of the 4-bus power system.

As shown in Fig. 9, a 4-bus test system modified from that in [39] is first used to demonstrate the proposed VSA approach. It has an external grid supporting three constant power loads. The external grid model is provided by the software DIgSILENT PowerFactory [40]. The time-domain simulation is executed by DIgSILENT PowerFactory with gradual load increases from the original operating condition. Then the proposed HEM-based VSA is implemented in MATLAB, and compared with the CPF using MATPOWER 6.0 toolbox [41]. Case A and Case B are quasi-dynamic simulations considering different types of load profiles. Case C is a full dynamic simulation with a generator connected to Bus 4, shown in Fig. 14. ZIP models for the loads as well as control models for the generator are also integrated. The technical data for Case A, B and C are given in the APPENDIX-C.

### *A. Case A: Loads increasing at the same rate*

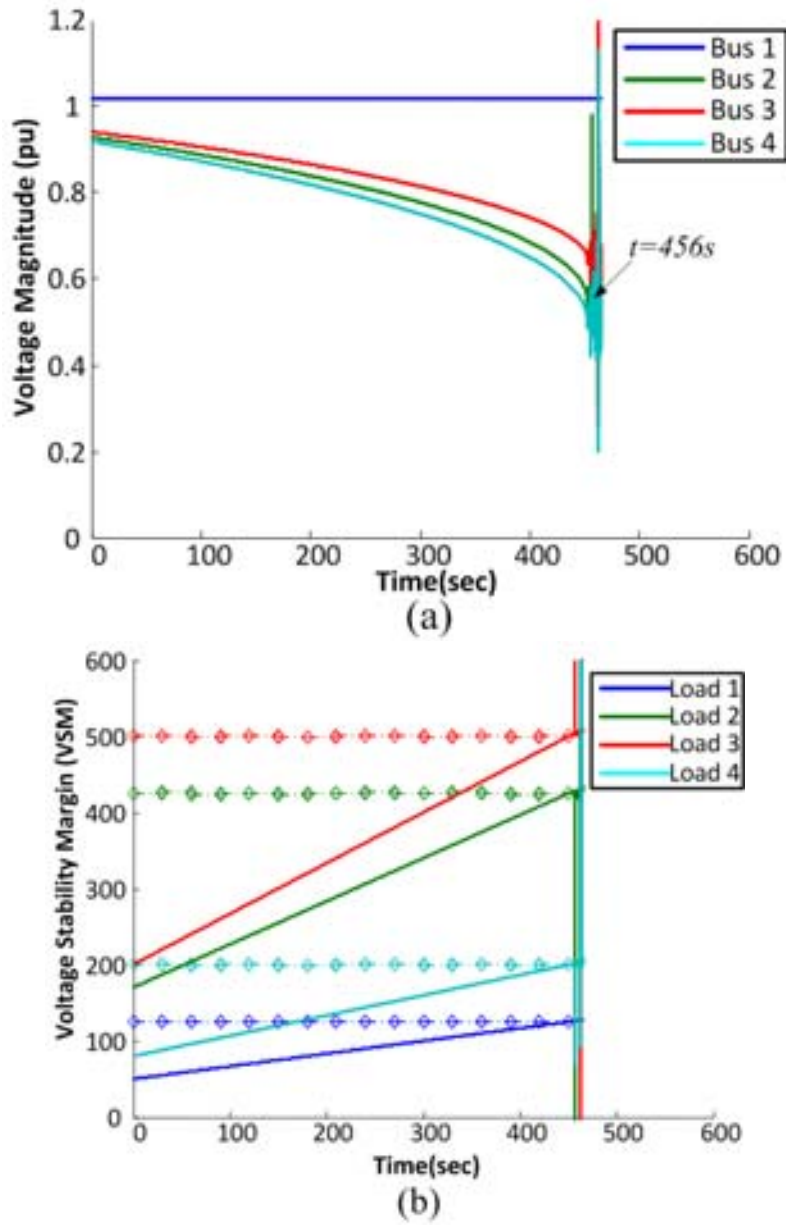

Fig. 10. (a) Voltage magnitudes (b) Active power of 4-bus system for Case A.

In Case A, the power consumption of all loads is gradually ramped up at the same rate from their original loads while their power factors are kept unchanged. It can be identified by the voltage magnitudes in Fig. 10(a) that voltage collapse happens at $t$ = 456 s, which is accurately predicted by the results from the HEM calculated every 30 seconds, as shown in Fig. 10(b), where active powers at all buses cross their respective voltage stability limits at the same time. The diamond markers connected to horizontal dash-dot lines in Fig. 10(b) indicate the voltage limit of each bus calculated by the HEM and updated every 30 seconds (assumed to be the time interval of state estimation). Notice from the Fig. 10(b) that this case has invariant voltage stability limits for all buses, since all loads increase in the same pattern.

### *B. Case B: Loads increasing at different random rates with variable power factors*

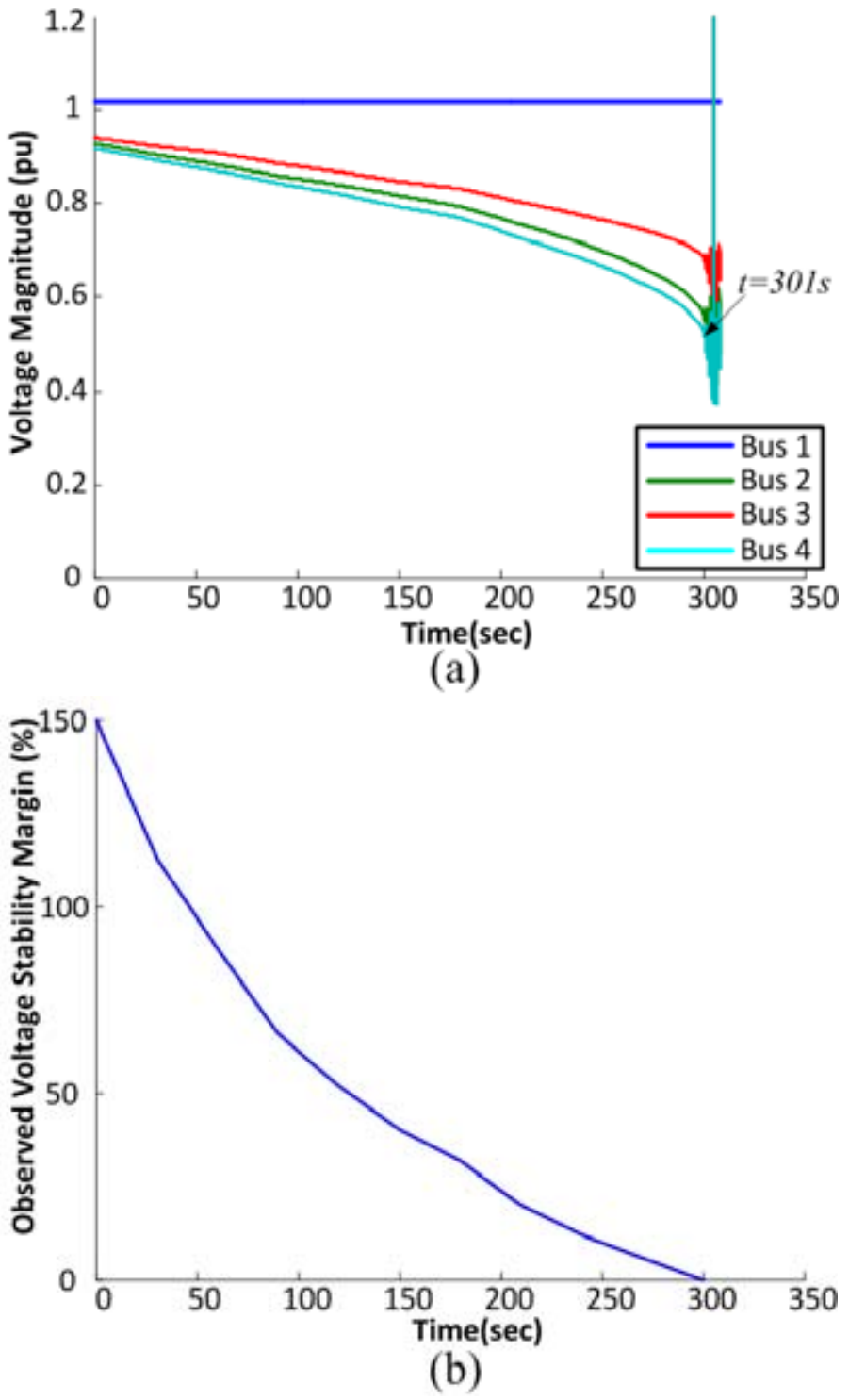

Fig. 11. (a) Voltage magnitudes (b) VSM of 4-bus system for Case B.

In the time-domain simulation of Case B, the active power and reactive power of all loads are gradually increased every 30 seconds at random rates, ranging between 0% and 30% of their original loads. Seen from voltage magnitudes of the 4-bus system in Fig. 11(a), voltage collapse happens at $t$ = 301 s. This voltage collapse is also accurately predicted by the overall VSM calculated by (28) from the HEM in Fig. 11(b). Fig 12 compares the active and reactive powers of each load to the voltage stability limits, which are shown by the dash-dot lines and updated every 30 seconds based on the result of state estimation and load trending prediction. Also, note that, in Fig. 12(a), the active powers of loads at Buses 2, 3 and 4 cross their limits at different times, i.e. $t$ = 290 s, 295 s and 289 s, respectively. In Fig. 12(b), the reactive power of load at Bus 4 crosses the limit at $t$ = 290 s.

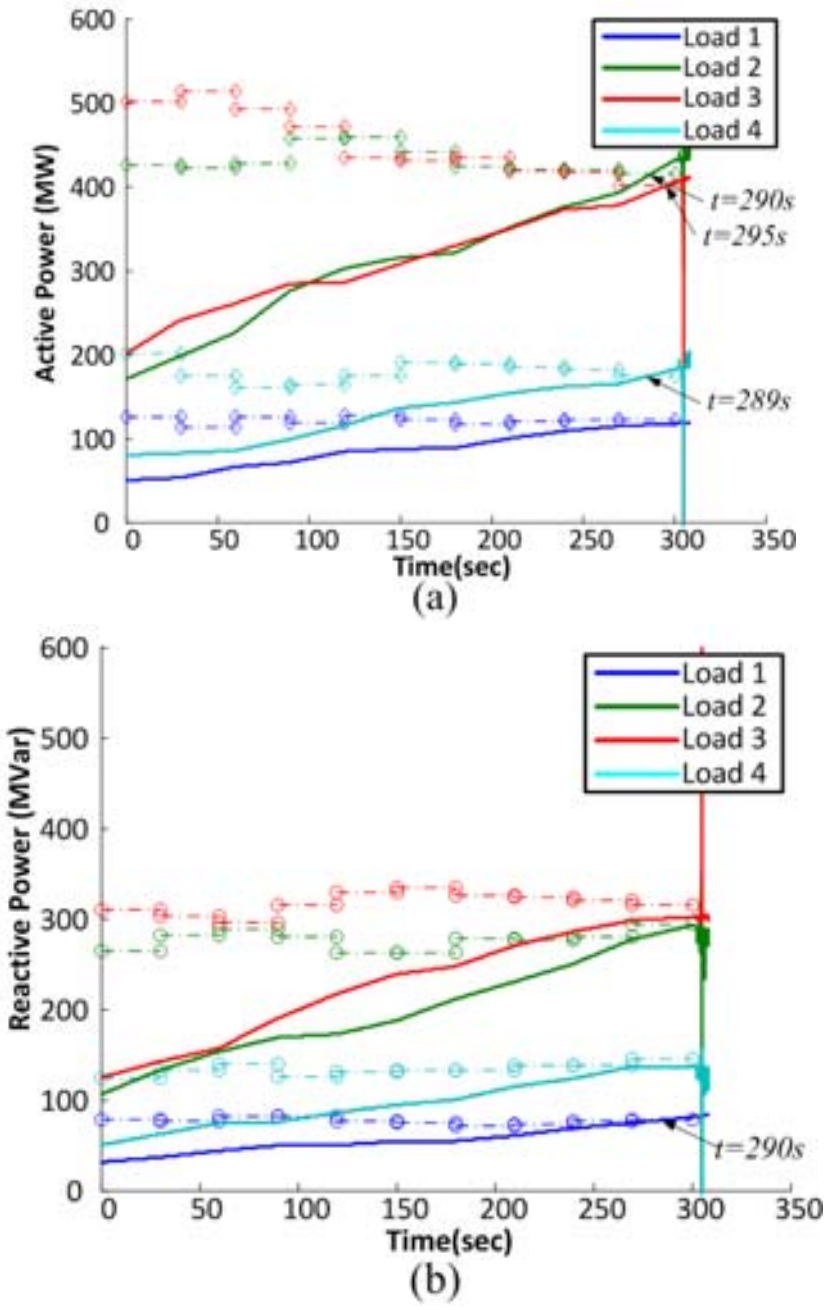

Fig. 12. Powers and stability limits of 4-bus system for Case B (a) active powers and limits (b) reactive power and limits.

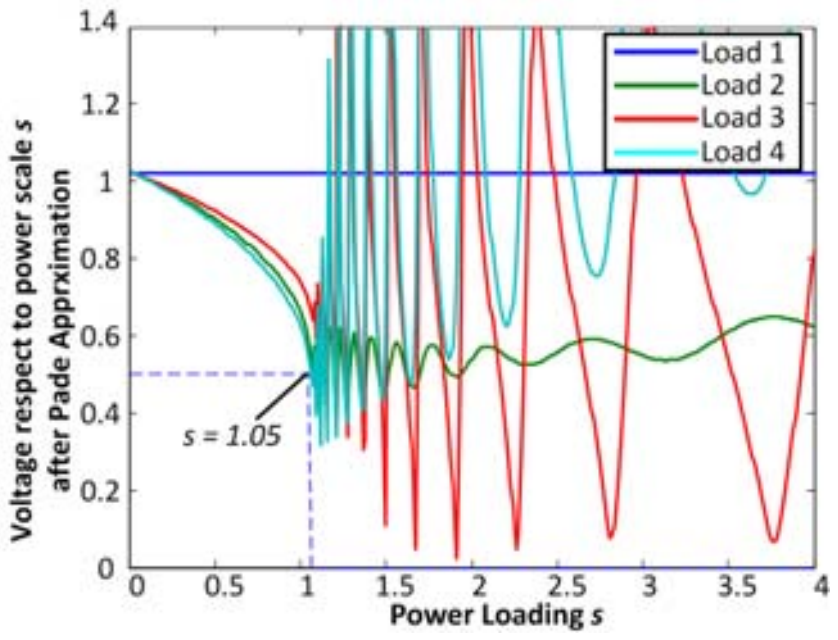

Fig. 13. The voltage w. r. t. loading scale predicted by HEM and Pade approximants at $t$ = 240 s.

Fig. 13 shows the voltage with respect to the overall loading scale predicted by the HEM at $t$ = 240 s. This collapse loading scale $sc$ is 1.05, so the VSM is only 5%, which means that if the loads in the next 30 seconds ramp up at an unchanged rate, the total load can only increase by 5% at the end of the next period, i.e. at $t$ = 270 s. Seen from the screening result in Fig. 13, which is implemented in MATLAB using the HEM at $t$ = 240 s, Bus 4 is the critical bus having the lowest voltage, where remedial control actions should be taken first.

### *C. Case C: Loads increasing at different random rates, considering ZIP load and dynamic models for the generator*

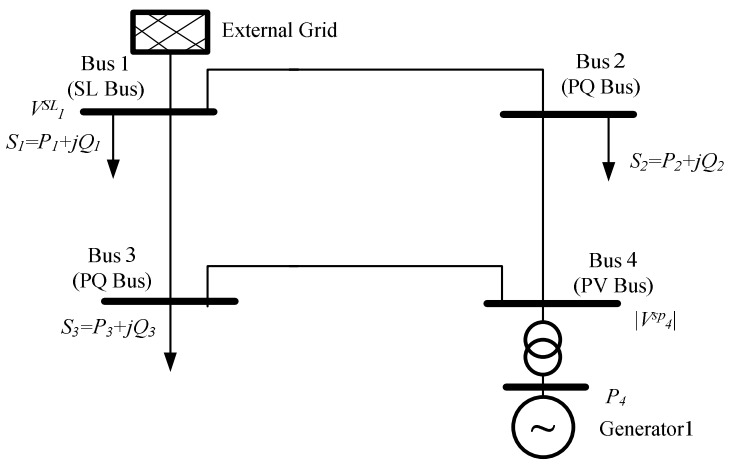

Fig. 14. One-line diagram of the 4-bus power system with a generator bus.

As shown in Fig. 14, Bus 4 is connected to a generator with 238MW active power output. The generator is with 6th order detailed model and equipped with IEEE Type 3 speed-governor model and IEEE Type 1 excitation model with over excitation limit. During the simulation, around 20% of the total active power load increase is picked up by the generator, while the remaining is balanced by the external grid, i.e. $\alpha_i$ = 20% in (25). ZIP load model is considered with the percentages of constant Z, constant I and constant P as 10%, 30% and 60% respectively.

The active power and reactive power of all loads are gradually increased at random rates, the same as Case B. Seen from voltage magnitudes of the time-domain simulation in Fig. 15(a), with voltage support from the generator at Bus 4, voltage collapse happens at $t$ = 506 s, which is later than Case B. Fig. 15(b) shows the active and reactive power output of the generator, where the over excitation limit is activated at $t$ = 357 s.

This voltage collapse can be predicted by the overall VSM from the HEM in Fig. 16. The VSM approaches 0 at $t$ = 480 s. Fig. 17 compares the active and reactive powers of each load to the voltage stability limits. In Fig. 17(a), the active loads at buses 1, 2 and 3 cross their respective limits at the same time with the update of state estimation, i.e. $t$ = 480 s. In Fig. 17(b), the reactive loads at buses 2 and 3 cross the limits at different time, i.e. $t$ = 480 s and 496 s. Fig. 18 shows the voltage with respect to the overall loading scale screened by the HEM at $t$ = 480 s. It can be noticed that voltage at Bus 3 has the lowest voltage and is thus identified as the critical bus where remedial actions should be taken first.

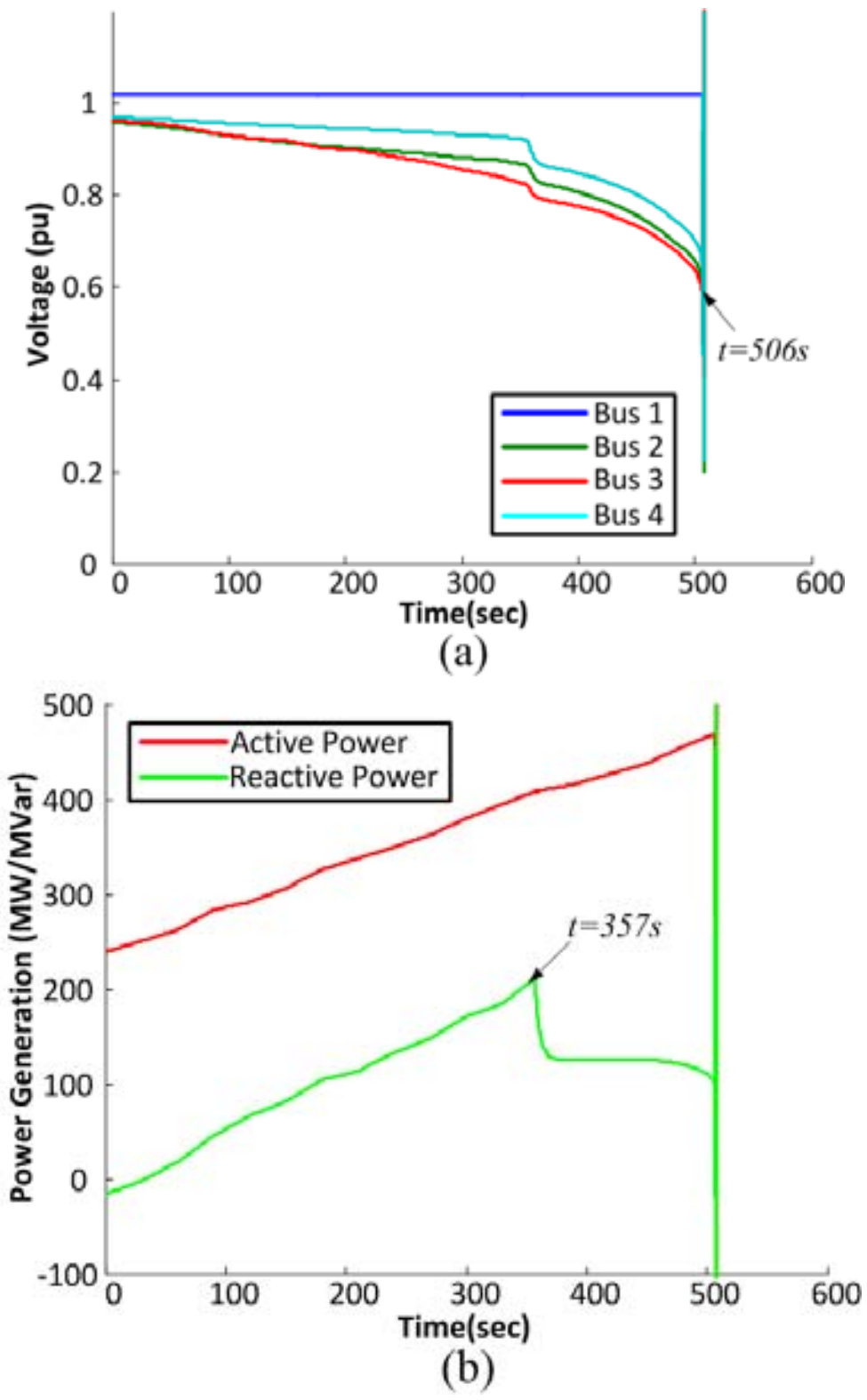

Fig. 15. (a) Voltage magnitudes (b) Generator's power output of the 4-bus system for Case C.

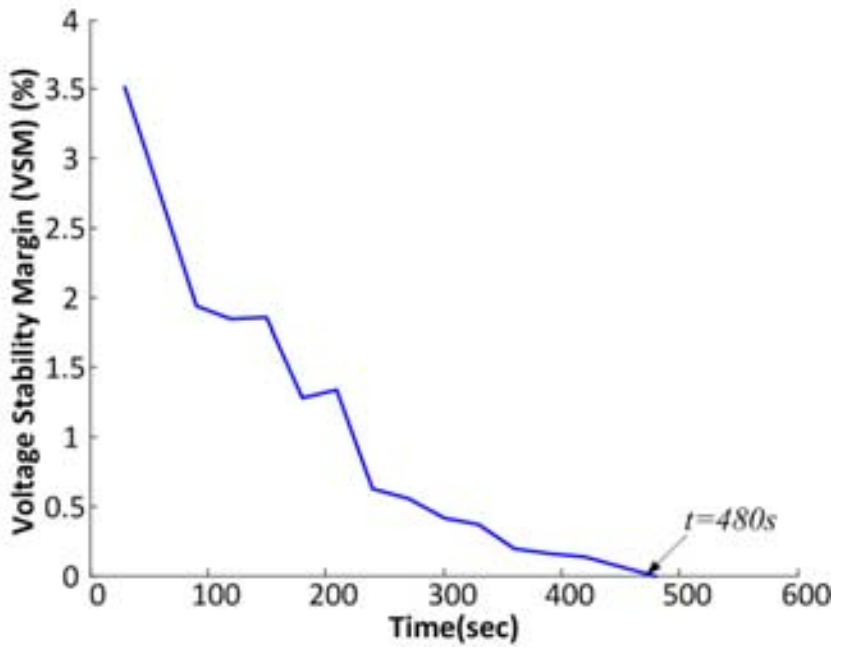

Fig. 16. The VSM of 4-bus system for Case C.

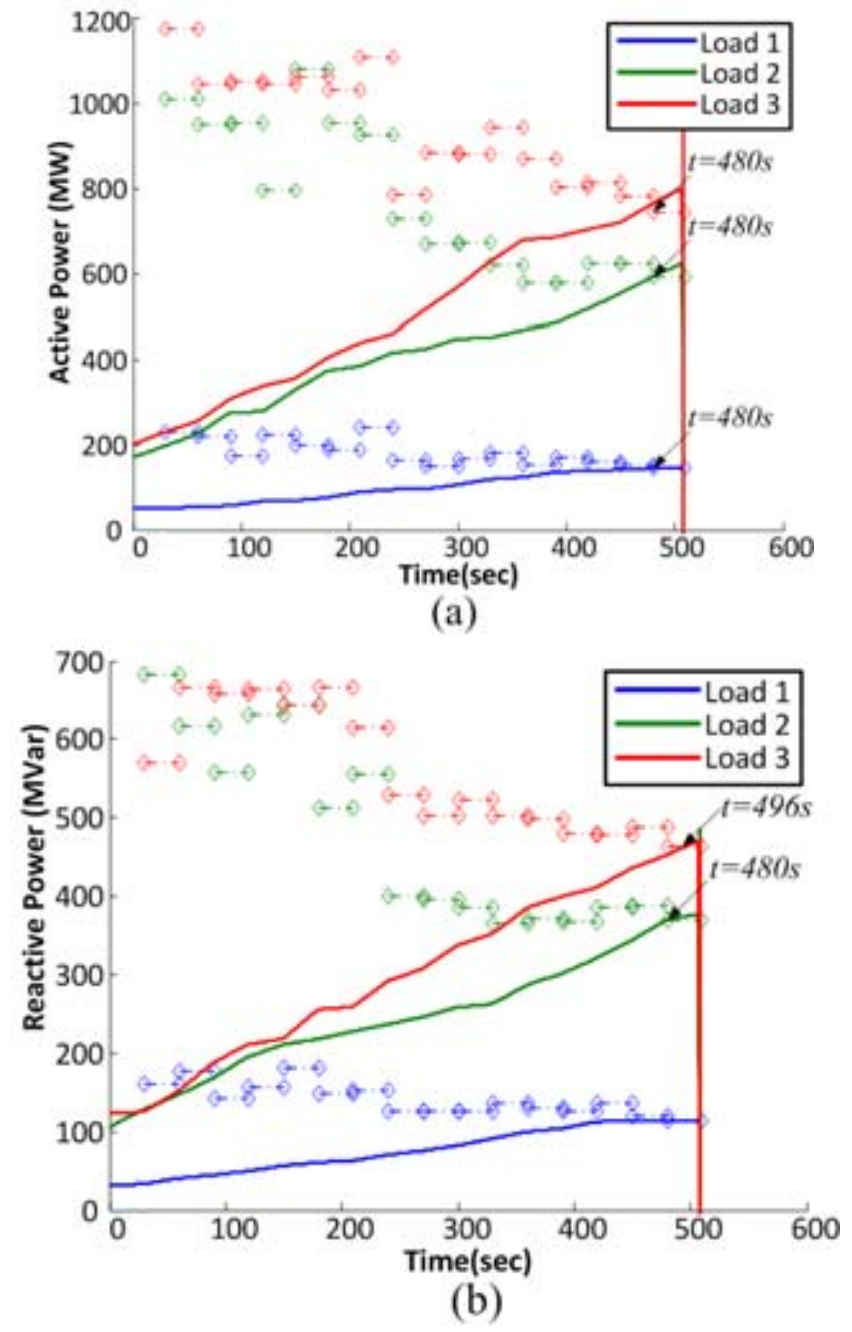

Fig. 17. Active and reactive powers and respective stability limits for Case C (a) active powers and limits (b) reactive powers and limits.

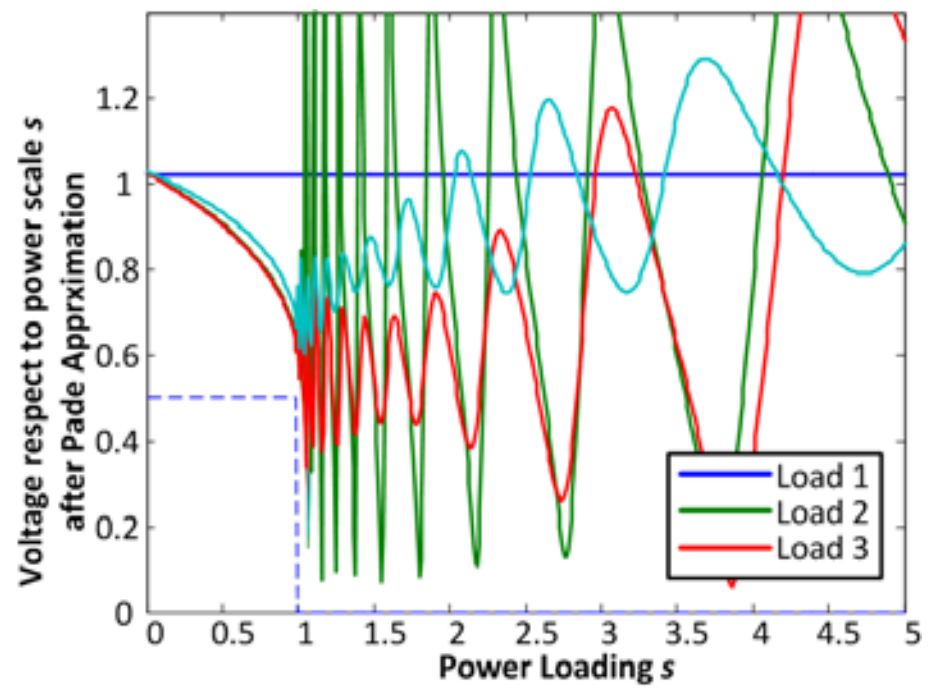

Fig. 18. The voltage w. r. t. loading scale predicted by HEM and Pade approximants at $t$ = 480 s for Case C.

In order to show the advantage of the proposed method for VSA, the computation time is compared between the proposed HEM-based VSA and the CPF. Table II shows the maximum computation time among all scenarios using the proposed method and the CPF. The CPF is carried out in MATPOWER 6.0 adopting natural parameterization and adaptive step length control. Both algorithms are executed on a PC with Intel Core i5-3210M 2.50GHz CPU and 4GB RAM and both algorithms are configured with an error tolerance of 10e-6. It can be noted that the CPF has a number of prediction-correction steps $N_{Step}$ and its maximum computation time among all scenarios is between 11.10 s and 20.64 s, while the proposed method is much faster, with maximum computation time between 0.873s and 1.473s. This can be simply explained by the non-iterative nature of the HEM. For a single loading level with the same precision, the numbers of terms ($N_{Germ}$ and $N_{PS}$) to be calculated by the proposed HEM-based VSA scheme are smaller than the steps of the CPF ($N_{Step}$), and each term is calculated by matrix multiplication instead of the iterative computations of the Newton-Raphson method for each step of the CPF.

TABLE II.
COMPARISON OF MAXIMUM COMPUTATION TIME BETWEEN THE PROPOSED METHOD AND THE CONTINUATION POWER FLOW METHOD

| Method | HEM-based VSA | | | | | | CPF | |
|---|---|---|---|---|---|---|---|---|
| Max.term /Time (s) | $N_{Germ}$ | $N_{PS}$ | $T_{Germ}$ | $T_{PS}$ | $T_{Pade}$ | $T_{HEM}$ | $N_{Step}$ | $T_{CPF}$ |
| **Case A** | 6 | 14 | 0.003s | 0.55s | 0.32s | **0.873s** | 232 | **11.10s** |
| **Case B** | 6 | 15 | 0.003s | 0.64s | 0.65s | **1.293s** | 238 | **13.33s** |
| **Case C** | 6 | 29 | 0.003s | 0.9s | 0.57s | **1.473s** | 266 | **20.64s** |

## VI. CASE STUDY ON THE NPCC TEST SYSTEM

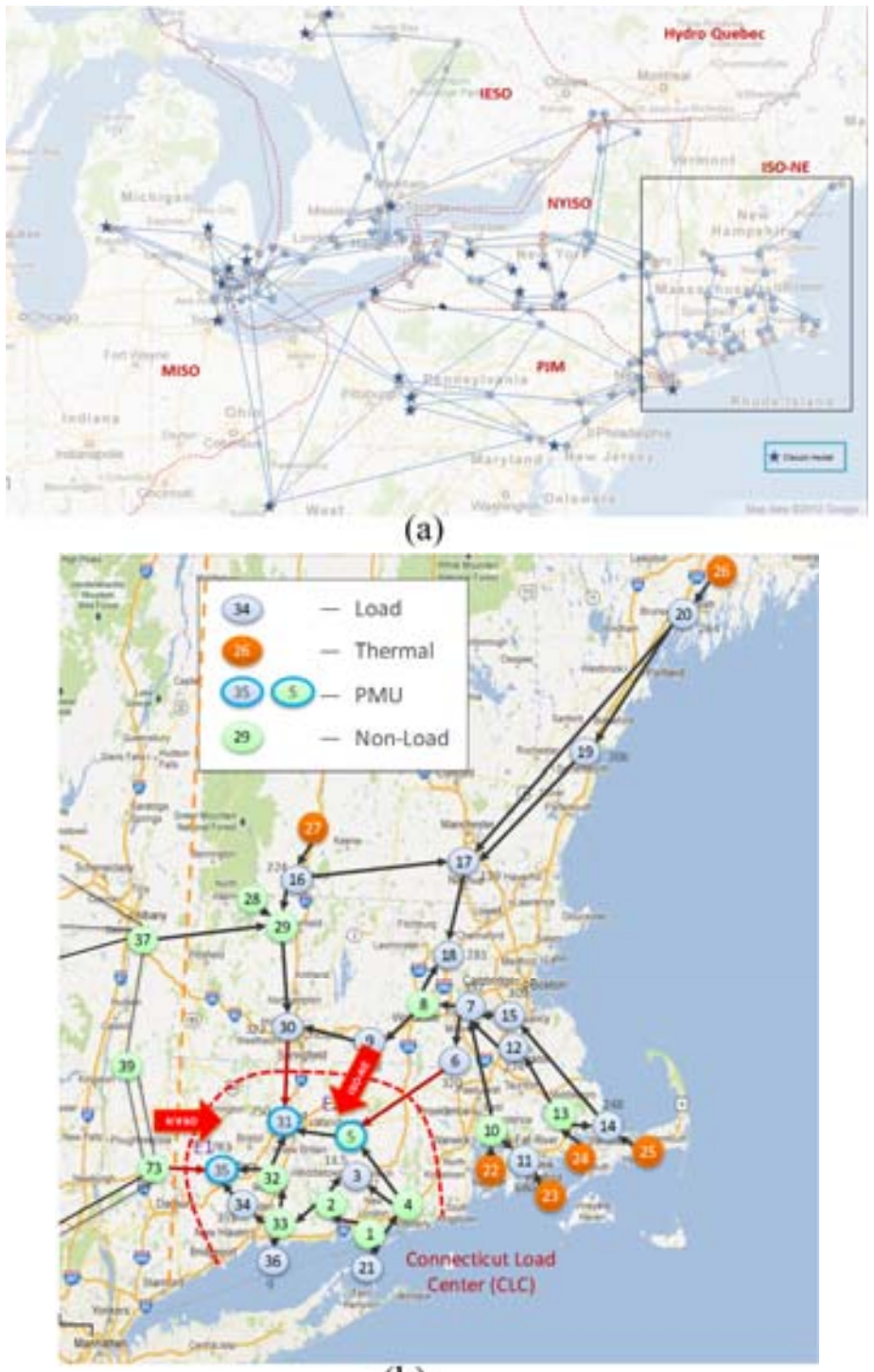

(a)

(b)

Fig. 19. Map of NPCC system and the CLC load area (a) system topology (b) CLC area.

The NPCC system with 48 generators and 140 buses in [42] is adopted to demonstrate the proposed HEM-based VSA method. As shown in Fig. 19, the Connectiut Load Center (CLC) area is supported by three tie lines, i.e. 73-35, 30-31 and 6-5. Assume that all 3 tie lines are equipped with PMUs at boundary buses 35, 31 and 5 of the load area. PowerFactory is used to simulate the voltage instability scenarios in the following cases. The load at each bus of the area is modelled as constant power load. All buses in load area are measured by SCADA system and the state estimation results on the load area are also assumed to be updated every 30 seconds. Thus, the HEM-based VSA scheme aims at predicting voltage instability in the next 30 seconds period.

### A. Case D: Load increasing at same rate

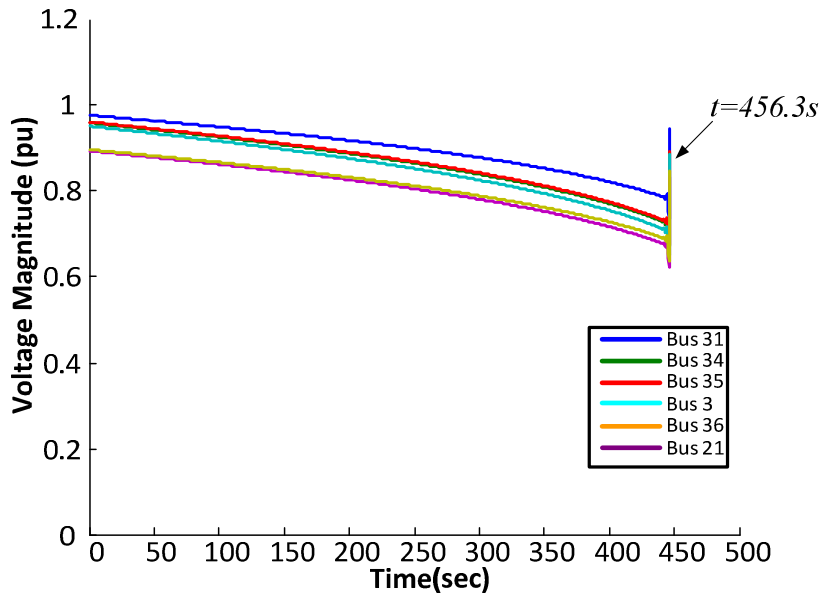

Fig. 20. Voltage magnitudes for Case D.

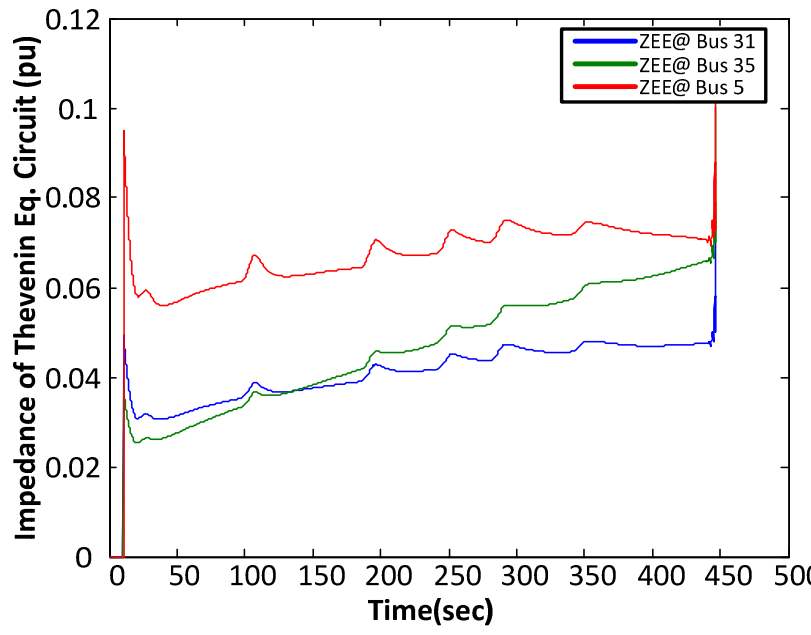

Fig. 21. Estimation of the impedance of equivalent branches for Case D.

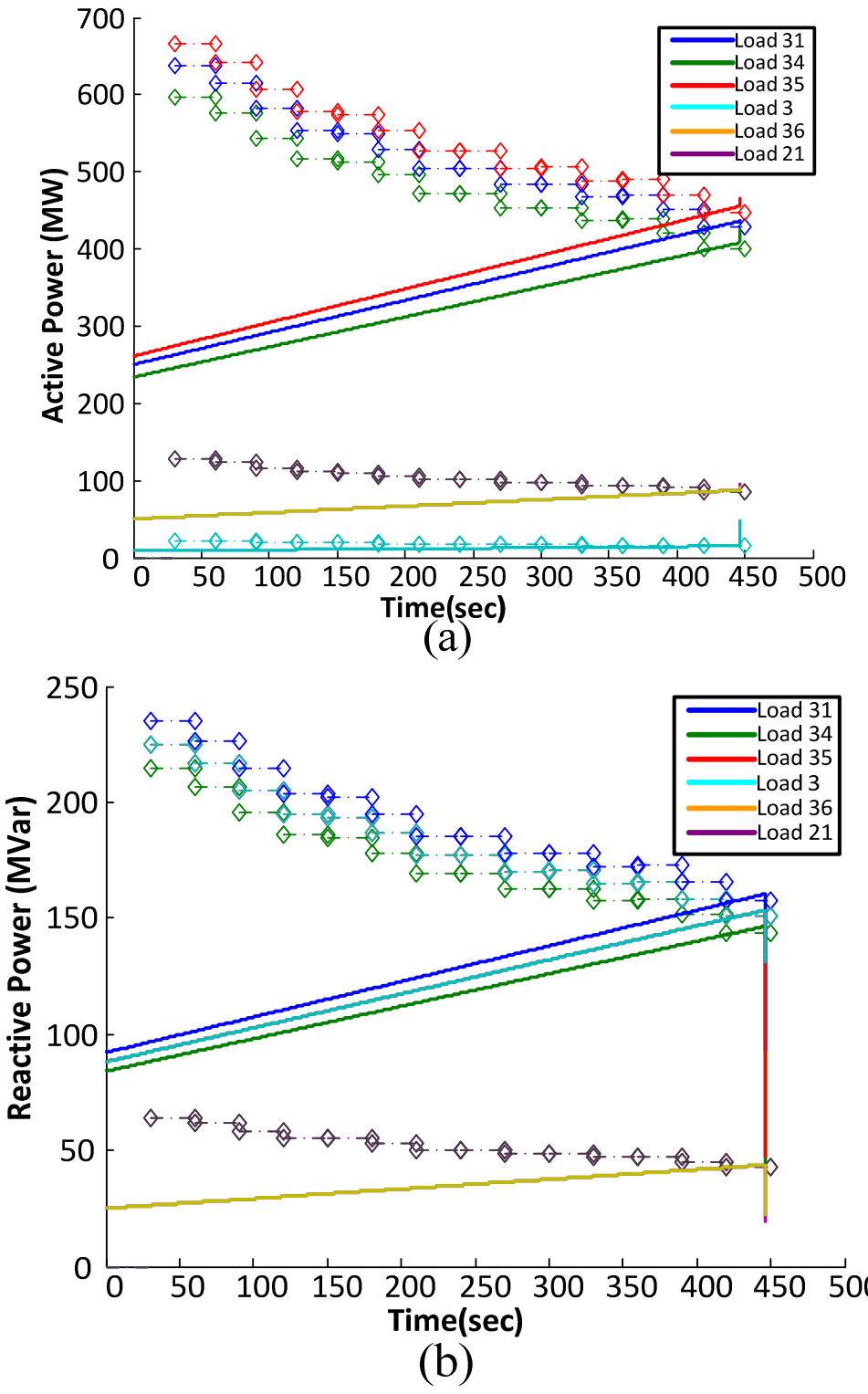

Fig. 22. Active and reactive powers and respective stability limits for Case D (a) active powers and limits (b) reactive powers and limits.

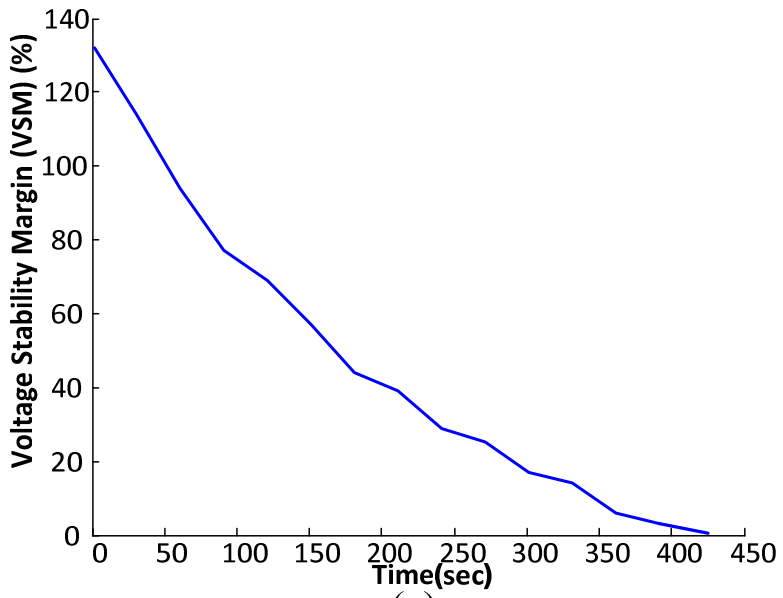

(a)

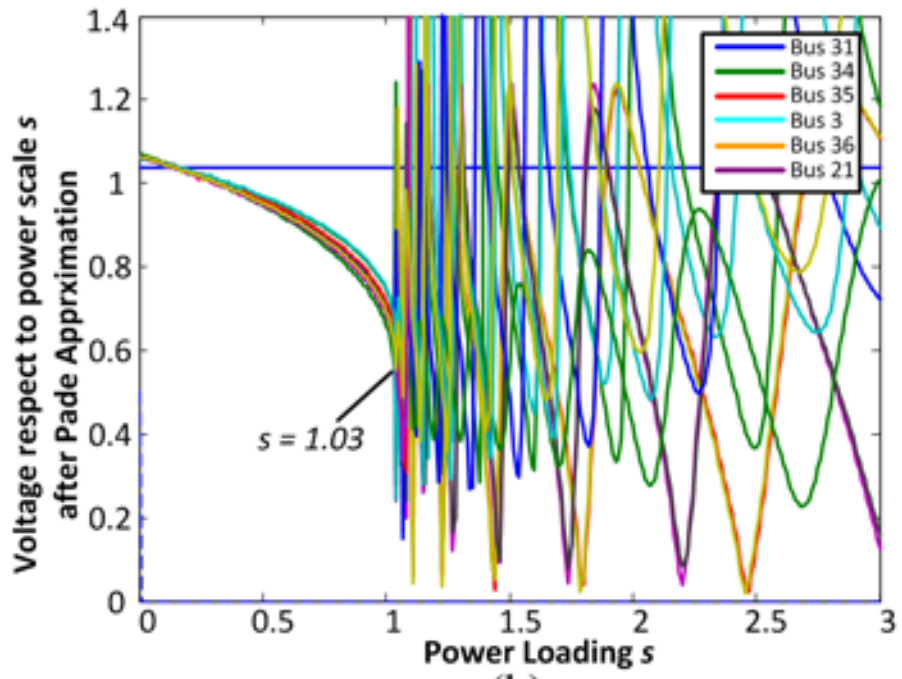

(b)

Fig. 23. (a)VSM (b) Voltage w. r. t. loading scale predicted by the HEM and Pade approximants at $t = 420$ s for Case D.

In the dynamic simulation of Case D, loads at all buses of the area increase at a same rate of 0.16% per second. Fig. 20 shows that voltages gradually decrease and then collapse at $t = 456.3$ s. The external system is aggregated to an equivalent voltage source $E$ directly connected to three boundary buses of the load area by three branches. Fig. 21 gives the estimates of the branch impedance to buses 31, 35 and 5. Voltage collapse is predicted when the active and reactive powers meeting their limits during $t = 420$-$450$ s, as shown by dash-dot lines in Fig. 22. Compared to Case A of the 4-bus system, the variations in stability limits are caused by the changes in the external system and consequently in their equivalent parameters. As shown in Fig. 23(a), at $t = 420$ s the VSM decreases to only 3%. Assuming a threshold of control activation is preset, e.g. 10%, the action should have been taken, e.g. switching in a shunt capacitor at the critical bus having the lowest voltage, i.e. Bus 34. Fig. 23(b) gives the voltage with respect to loading scale $s$ predicted by the HEM at $t = 420$ s, and a precise collapse load scale $sc = 1.03$.

### *B. Case E: Loads increasing at different random rates with variable power factors*

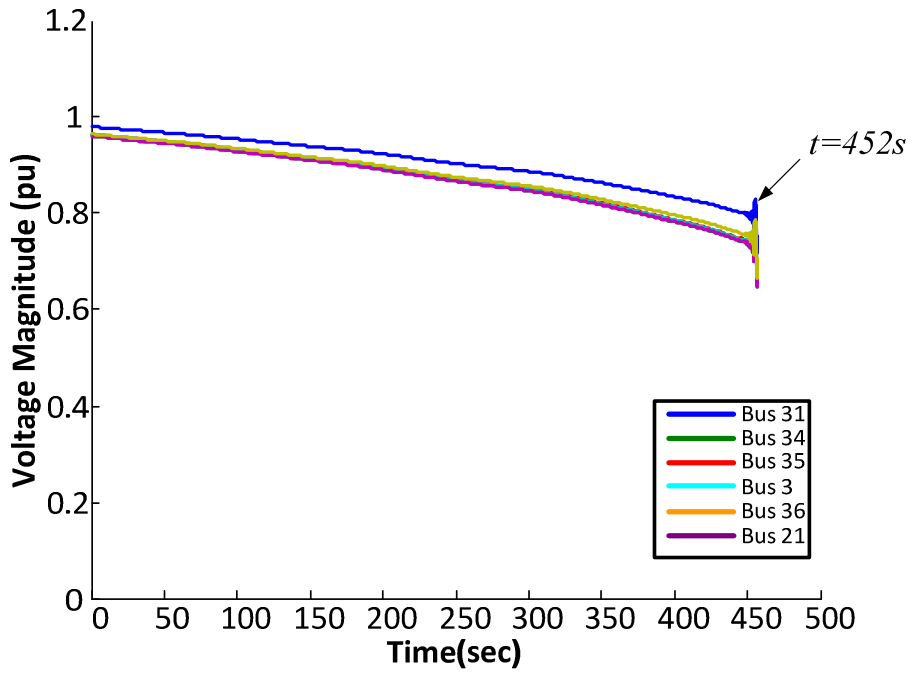

Fig. 24. Voltage magnitudes for Case E.

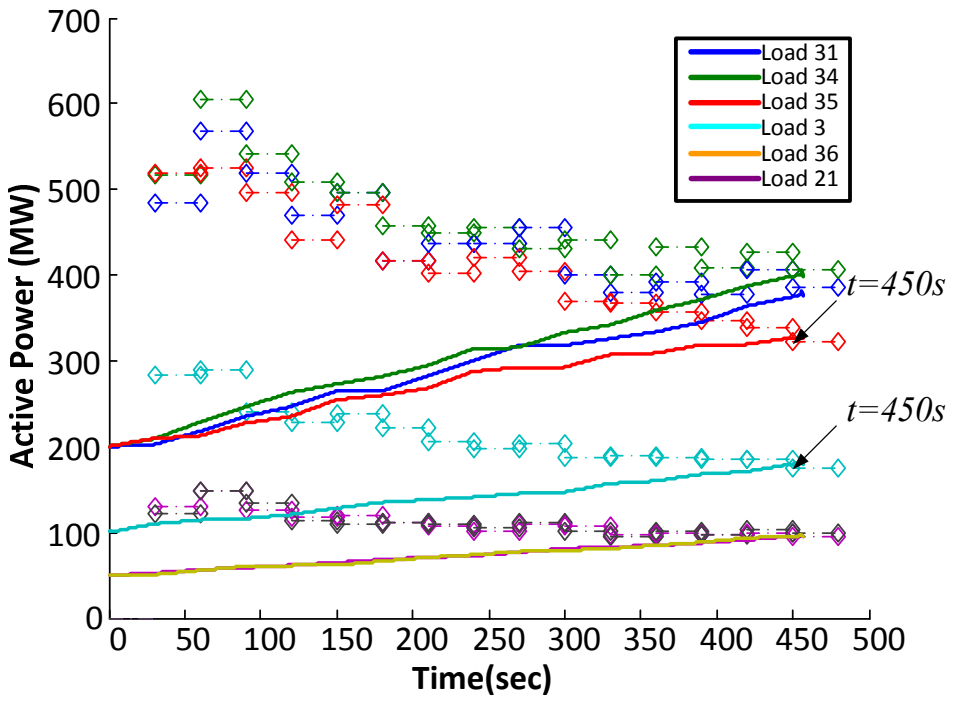

(a)

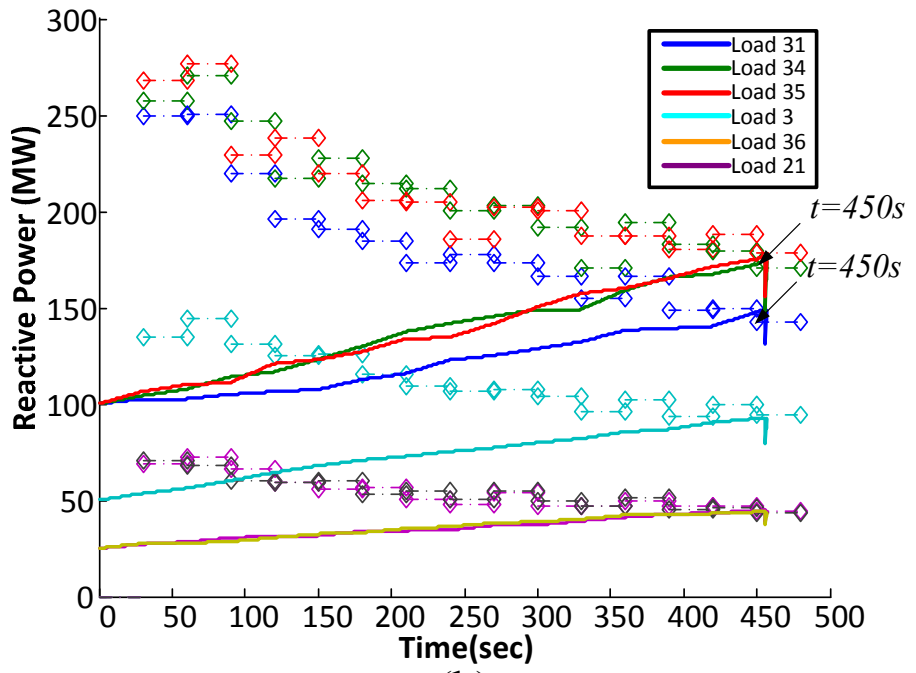

(b)

Fig. 25. Active and reactive powers and respective stability limits for Case E (a) active powers and limits (b) reactive powers and limits.

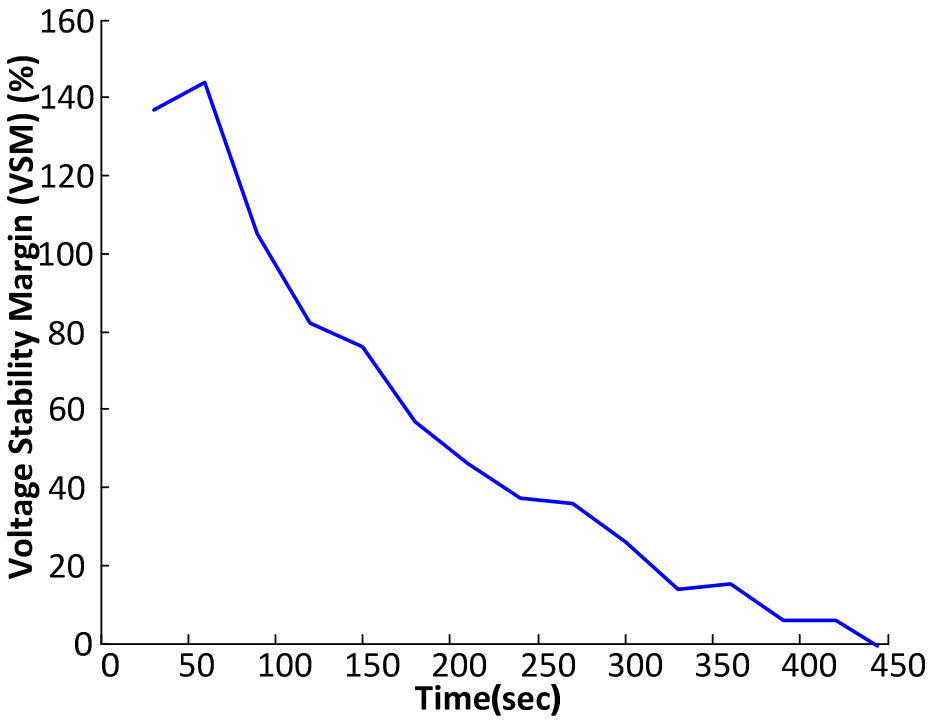

Fig. 26. VSM predicted by the HEM for Case E.

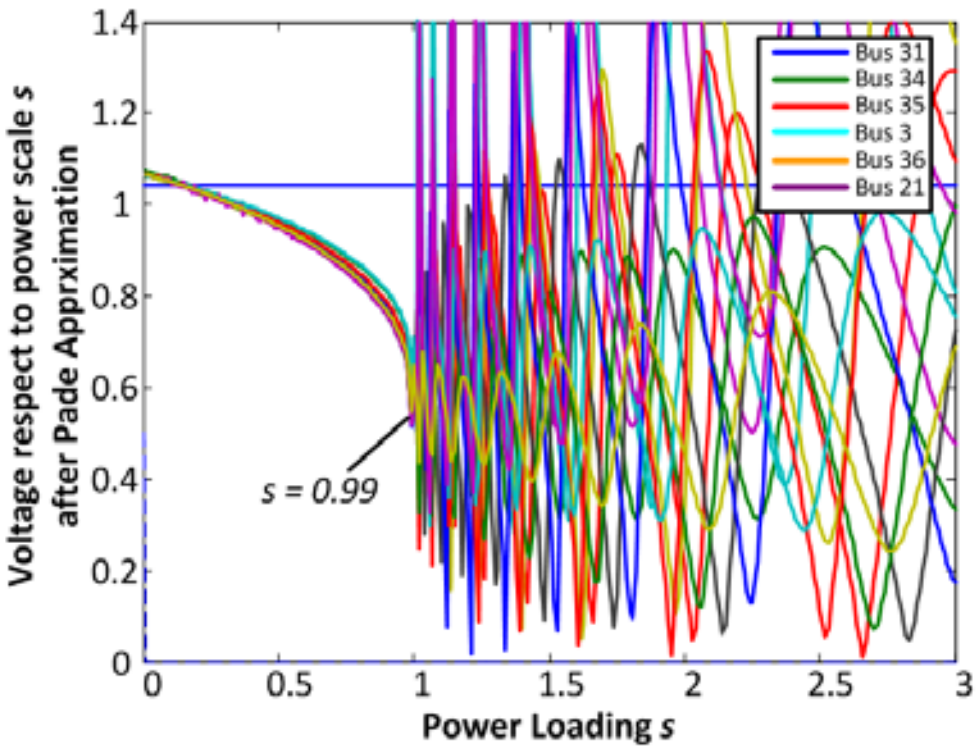

Fig. 27. Voltage w. r. t. loading scale predicted by HEM and Pade approximants at $t = 450$ s for Case E.

In this case, a more realistic scenario with loads increasing at a random rate in 0-10% every 30 seconds is tested. Voltage collapse happens at $t = 452$ s, as shown in Fig. 24. The last warning should be armed between $t$ = 420 s and $t$ =450 s with a more conservative threshold, since loads at Bus 31, 34, 35 and 3 cross their individual stability limits at $t$ = 450 s, as shown in Fig. 25 (a) and (b) for active power and reactive power respectively. The VSM is below 0 updated at $t$ = 450 s, as shown in Fig. 26, indicating the occurrence of voltage instability. Fig. 27 shows the voltage with respect to loading scale $s$ predicted by the HEM at $t$ = 450 s, in which the precise collapse load scale $sc$ = 0.99 as a precursor signal of immediate voltage collapse.

The online HEM-based VSA scheme is able to provide system operators with a timely and accurate indication of the VSM. Practically, a more conservative threshold of VSMs should be preset, e.g. 5-10%. On one hand, that gives operators enough time to activate emergency voltage control; on the other hand, it leaves enough stability margin to prevent voltage collapse. All computations of this new scheme can be finished within 2 seconds in MATLAB, including external parameter identification, load flow trending prediction, the proposed HEM, and adaptive Pade approximants, so the overall VSM and individual VSM for each bus is predicted about 28 seconds ahead. If each bus is equipped with a PMU to provide synchronized phasor data at a higher frequency, e.g. 30-60Hz, operators can monitor the load change and the VSM at each load bus with higher accuracy so as to decide more precisely whether and when a remedial action is needed.

Table III compares the maximum computation time among all scenarios in Case D and Case E using the proposed method and the CPF with the same error tolerance, i.e. 10e-6. It can be noted that the proposed HEM-based VSA scheme has better time performance.

TABLE III.
COMPARISON OF MAXIMUM COMPUTATION TIME BETWEEN THE PROPOSED METHOD AND THE CONTINUATION POWER FLOW METHOD

| Method | HEM-based VSA | | | | | | CPF | |
|---|---|---|---|---|---|---|---|---|
| Max.term /Time (s) | $N_{Germ}$ | $N_{PS}$ | $T_{Germ}$ | $T_{PS}$ | $T_{Pade}$ | $\mathbf{T_{HEM}}$ | $N_{Step}$ | $\mathbf{T_{CPF}}$ |
| **Case D** | 8 | 21 | 0.005s | 0.9s | 0.52s | **1.425s** | 228 | **14.10s** |
| **Case E** | 8 | 23 | 0.005s | 1.1s | 0.63s | **1.735s** | 235 | **16.36s** |

## VII. CONCLUSION

This paper proposes a new online VSA scheme for a load area based on the proposed HEM with a physical germ solution. The physical germ solution endows the physical meaning of the analytical expression, i.e. the loading level. Based on the result from the HEM, it can quickly monitor the VSM of a load area, predict the voltage instability and suggest system operators the timing and location to take remedial actions. An adaptive two-stage Pade approximants method is also proposed to extend the convergence region of the power series, making the prediction of voltage collapse more accurate. These methods jointly contribute to the performance of the proposed VSA. This new VSA scheme has been demonstrated on the 4-bus power system and on the 140-bus NPCC system.

Compared with model-based VSA methods, the proposed scheme has better time performance by using the non-iterative HEM to give the whole P-V curve. Compared with measurement-based methods, the proposed scheme avoids network reduction in the load area while keeping acceptable online performance. Also, this scheme can integrate load trending prediction at each bus to provide look-ahead voltage stability monitoring.

## APPENDIX A

### *Finding Physical Germ Solution of a 3-Bus System*

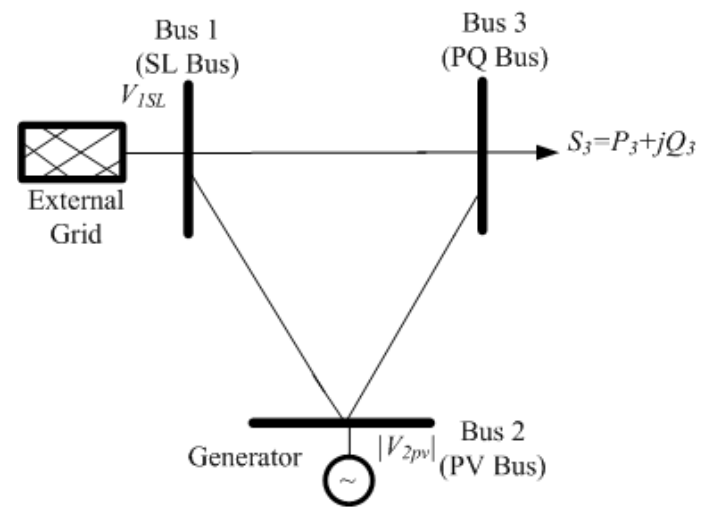

Fig. 28. Diagram of the demonstrative 3-bus system.

A 3-bus system having 1 slack bus, 1 PV bus and 1 PQ bus, shown in Fig. 28, is adopted to demonstrate the procedure of finding the physical germ solution. The first step is to calculate the starting no-load no-generation condition. Only the slack bus propagates its voltage to the whole passive network (A1).

$$\begin{bmatrix} 1 & & & & & \\ & 1 & & & & \\ G_{21} & -B_{21} & G_{22} & -B_{22} & G_{23} & -B_{23} \\ B_{21} & G_{21} & B_{22} & G_{22} & B_{23} & G_{23} \\ G_{31} & -B_{31} & G_{32} & -B_{32} & G_{33} & -B_{33} \\ B_{31} & G_{31} & B_{32} & G_{32} & B_{33} & -B_{33} \end{bmatrix} \begin{bmatrix} V_{ST1re} \\ V_{ST1im} \\ V_{ST2re} \\ V_{ST2im} \\ V_{ST3re} \\ V_{ST3im} \end{bmatrix} = \begin{bmatrix} \mathrm{Re}\left(V_1^{SL}\right) \\ \mathrm{Im}\left(V_1^{SL}\right) \\ 0 \\ 0 \\ 0 \\ 0 \end{bmatrix} \quad \text{(A1)}$$

The second step is to find the physical germ solution by an embedding. The right hand side of the PQ bus (i.e. Bus 3) is 0, since no-load condition is held for the physical germ solution. Thus, the embedding of PV bus (i.e. Bus 2) is needed to adjust its voltage magnitude to the specified value $|V_{2pv}|$, while keep the active power generation as the specified value $P_2$.

Move all the $n^{th}$ order terms (i.e. the unknowns) to the left hand side and leave terms 0 to ($n$-1) (i.e. the knowns) on the right hand side of (A2). The matrix equation is extended by moving the unknown variables of PV bus (Bus 2), i.e. $W_{g2re}[n]$, $W_{g2im}[n]$ and $Q_{g2}[n]$ to the left hand side.

$$\begin{bmatrix} 1 & 0 & 0 & 0 & 0 & 0 & 0 & 0 & 0 \\ 0 & 1 & 0 & 0 & 0 & 0 & 0 & 0 & 0 \\ G_{21} & -B_{21} & G_{22} & -B_{22} & G_{23} & -B_{23} & 0 & 0 & W_{ST2im} \\ B_{21} & G_{21} & B_{22} & G_{22} & B_{23} & G_{23} & 0 & 0 & W_{ST2re} \\ G_{31} & -B_{31} & G_{32} & -B_{32} & G_{33} & -B_{33} & 0 & 0 & 0 \\ B_{31} & G_{31} & B_{32} & G_{32} & B_{33} & -G_{33} & 0 & 0 & 0 \\ 0 & 0 & W_{ST2re} & -W_{ST2im} & 0 & 0 & V_{ST2re} & -V_{ST2im} & 0 \\ 0 & 0 & W_{ST2im} & W_{ST2re} & 0 & 0 & V_{ST2im} & V_{ST2re} & 0 \\ 0 & 0 & V_{ST2re} & V_{ST2im} & 0 & 0 & 0 & 0 & 0 \end{bmatrix} \begin{bmatrix} V_{g1re}[n] \\ V_{g1im}[n] \\ V_{g2re}[n] \\ V_{g2im}[n] \\ V_{g3re}[n] \\ V_{g3im}[n] \\ W_{g2re}[n] \\ W_{g2im}[n] \\ Q_{g2}[n] \end{bmatrix} = \begin{bmatrix} 0 \\ 0 \\ \mathrm{Re}\left(P_2 W_{g2}[n-1] + \overset{n-1}{\underset{1}{Conv}}(Q_{g2} * W_{g2}^{*})\right) \\ \mathrm{Im}\left(P_2 W_{g2}[n-1] + \overset{n-1}{\underset{1}{Conv}}(Q_{g2} * W_{g2}^{*})\right) \\ 0 \\ 0 \\ -\mathrm{Re}\left(\overset{n-1}{\underset{1}{Conv}}(W_{g2} * V_{g2})\right) \\ -\mathrm{Im}\left(\overset{n-1}{\underset{1}{Conv}}(W_{g2} * V_{g2})\right) \\ \varepsilon_2[n-1] \end{bmatrix}$$

(A2)

in which

$$\overset{n-1}{\underset{1}{Conv}}(Q_{g2} * W_{g2}^{*}) = \sum_{m=1}^{n-1} Q_{g2}[n-m] W_{g2}^{*}[m] \quad \text{(A3)}$$

$$\overset{n-1}{\underset{1}{Conv}}(W_{g2} * V_{g2}) = \sum_{m=1}^{n-1} W_{g2}[n-m] V_{g2}[m] \qquad \text{(A4)}$$

## APPENDIX B

### *HEM Considering ZIP Load for PQ Buses*

Consider a PQ bus $i$ with ZIP load given by (A5), where $p_Z$, $p_I$ and $p_P$ are the percentages of constant impedance, constant current and constant power components of active power load $P_i$, and $q_Z$, $q_I$ and $q_P$ are the percentages of constant impedance, constant current and constant power components of reactive power load $Q_i$. $P_{i0}$ and $Q_{i0}$ are the active and reactive powers at the nominal voltage 1.0pu, respectively. The PFE for ZIP load bus becomes (A6), which develops to (A7).

$$\begin{cases} P_i = P_{i0}(p_Z |V_i|^2 + p_I |V_i| + p_P) \\ Q_i = Q_{i0}(q_Z |V_i|^2 + q_I |V_i| + q_P) \end{cases}, \quad \forall i \in \mathcal{P} \qquad \text{(A5)}$$

$$\sum_{k=1}^{N} Y_{ik} V_k(s) = \frac{s\left(P_{i0}(p_Z |V_i|^2 + p_I |V_i| + p_P) - jQ_{i0}(q_Z |V_i|^2 + q_I |V_i| + q_P)\right)}{V_i^*(s^*)} \qquad \text{(A6)}$$

$$\begin{aligned} &\sum_{k=1}^{N} Y_{ik}\left(V_k[0] + V_k[1]s + V_k[2]s^2 + \cdots\right) \\ &= s\left\{(P_{i0}p_Z - jQ_{i0}q_Z)V_i(s) + \frac{(P_{i0}p_P - jQ_{i0}q_P) + (P_{i0}p_I - jQ_{i0}q_I)|V_i|(s)}{V_i^*(s^*)}\right\} \\ &= s\left\{S_{Zi}^* V_i(s) + \frac{S_{Pi}^* + S_{Ii}^* |V_i|(s)}{V_i^*(s^*)}\right\} \\ &= sS_{Zi}^*\left(V_i[0] + V_i[1]s + \cdots\right) + sS_{Pi}^*\left(W_{ig}^* + W_i^*[1]s + \cdots\right) \\ &+ sS_{Ii}^*\left(M_i[0] + M_i[1]s + \cdots\right)\left(W_{ig}^* + W_i^*[1]s + \cdots\right) \end{aligned} \qquad \text{(A7)}$$

where $S_{Zi}$, $S_{Ii}$, $S_{Pi}$ are the complex power of the load for constant impedance, constant current and constant power components respectively at bus $i$ and $M_i(s)$ is the series of voltage magnitude at bus $i$, i.e. $M_i(s) = |V_i(s)|$, satisfying (A9)

$$\begin{cases} S_{Zi} = P_{i0}p_Z + jQ_{i0}q_Z \\ S_{Ii} = P_{i0}p_I + jQ_{i0}q_I \\ S_{Pi} = P_{i0}p_P + jQ_{i0}q_P \end{cases} \qquad \text{(A8)}$$

$$M_i(s) \cdot M_i(s) = V_i(s) \cdot V_i^*(s^*) \qquad \text{(A9)}$$

Finally, if the ZIP load model is considered, the final recursive matrix equation of the demonstrative 3-bus system in Appendix A finally becomes (A11), in which

$$\overset{n-1}{\underset{0}{Conv}}(M_3 * W_3^*) = \sum_{m=0}^{n-1} M_3[n-m] W_3^*[m] \qquad \text{(A10)}$$

## APPENDIX C

### *Technical Data of the 4-Bus System*

The technical data of the 4-bus system in Section V is given in TABLE IV for transmission lines, TABLE V and TABLE VI for buses with the original loads, and TABLE VII for the generator, respectively.

TABLE IV. TRANSMISSION LINE DATA FOR 4-BUS SYSTEM

| From Bus -To Bus | Resistance R (pu) | Reactance X (pu) | Shunt admittance Y (pu) |
|---|---|---|---|
| **1-2** | 0.01008 | 0.0504 | 0.1025 |
| **1-3** | 0.00744 | 0.0372 | 0.0775 |
| **2-4** | 0.00744 | 0.0372 | 0.0775 |
| **3-4** | 0.01272 | 0.0636 | 0.1275 |

TABLE V. BUS DATA FOR 4-BUS SYSTEM FOR CASE A, B

| Bus No. | Bus Type | $V_{sp}$(pu) | Generation | | Load | |
|---|---|---|---|---|---|---|
| | | | $P_G$(MW) | $Q_G$(MVar) | $P_L$(MW) | $Q_L$(MVar) |
| **1** | SL | 1.02 | - | - | 50 | 30.99 |
| **2** | PQ | - | 0 | 0 | 170 | 105.35 |
| **3** | PQ | - | 0 | 0 | 200 | 123.94 |
| **4** | PQ | - | 0 | 0 | 80 | 49.58 |

TABLE VI. BUS DATA FOR 4-BUS SYSTEM FOR CASE C

| Bus No. | Bus Type | $V_{sp}$(pu) | Generation | | Load | |
|---|---|---|---|---|---|---|
| | | | $P_G$(MW) | $Q_G$(MVar) | $P_L$(MW) | $Q_L$(MVar) |
| **1** | SL | 1.02 | - | - | 50 | 30.99 |
| **2** | PQ | - | 0 | 0 | 170 | 105.35 |
| **3** | PQ | - | 0 | 0 | 200 | 123.94 |
| **4** | PV | 0.98 | 238 | [-100, 200] | 0 | 0 |

TABLE VII. GENERATOR DATA FOR CASE C

| | | |
|---|---|---|
| $H$ | Inertia time constant (s) | 4.2 |
| $x_l$ | Leakage reactance (p.u.) | 0.125 |
| $X_d$, $X_q$ | Synchronous reactances (p.u.) | 1, 0.69 |
| $X_d'$, $X_q'$ | Transient reactances (p.u.) | 0.31, 0.5 |
| $X_d''$, $X_q''$ | Sub-transient reactances (p.u.) | 0.25, 0.25 |
| $T_{d0}'$, $T_{q0}'$ | Transient time constants (s) | 10.2, 0.41 |
| $T_{d0}''$, $T_{q0}''$ | Sub-transient time constants (s) | 0.05, 0.035 |

$$\begin{bmatrix} 1 & 0 & 0 & 0 & 0 & 0 & 0 & 0 & 0 \\ 0 & 1 & 0 & 0 & 0 & 0 & 0 & 0 & 0 \\ G_{21} & -B_{21} & G_{22} & -B_{22} & G_{23} & -B_{23} & -P_{g2} & Q_{g2} & W_{g2im} \\ B_{21} & G_{21} & B_{22} & G_{22} & B_{23} & G_{23} & Q_{g2} & P_{g2} & W_{g2re} \\ G_{31} & -B_{31} & G_{32} & -B_{32} & G_{33} & -B_{33} & 0 & 0 & 0 \\ B_{31} & G_{31} & B_{32} & G_{32} & B_{33} & -G_{33} & 0 & 0 & 0 \\ 0 & 0 & W_{g2re} & -W_{g2im} & 0 & 0 & V_{g2re} & -V_{g2im} & 0 \\ 0 & 0 & W_{g2im} & W_{g2re} & 0 & 0 & V_{g2im} & V_{g2re} & 0 \\ 0 & 0 & V_{g2re} & V_{g2im} & 0 & 0 & 0 & 0 & 0 \end{bmatrix} \begin{bmatrix} V_{1re}[n] \\ V_{1im}[n] \\ V_{2re}[n] \\ V_{2im}[n] \\ V_{3re}[n] \\ V_{3im}[n] \\ W_{2re}[n] \\ W_{2im}[n] \\ Q_2[n] \end{bmatrix} = \begin{bmatrix} 0 \\ 0 \\ \mathrm{Re}\left(\overset{n-1}{\underset{1}{Conv}}(Q_2 * W_2^*)\right) \\ \mathrm{Im}\left(\overset{n-1}{\underset{1}{Conv}}(Q_2 * W_2^*)\right) \\ \boxed{\mathrm{Re}\left(S_{3P}^* W_3^*[n-1] + S_{3Z}^* V_3[n-1] + S_{3I}^* \overset{n-1}{\underset{0}{Conv}}(M_3 * W_3^*)\right)} \\ \boxed{\mathrm{Im}\left(S_{3P}^* W_3^*[n-1] + S_{3Z}^* V_3[n-1] + S_{3I}^* \overset{n-1}{\underset{0}{Conv}}(M_3 * W_3^*)\right)} \\ -\mathrm{Re}\left(\overset{n-1}{\underset{1}{Conv}}(W_2 * V_2)\right) \\ -\mathrm{Im}\left(\overset{n-1}{\underset{1}{Conv}}(W_2 * V_2)\right) \\ \delta_{n1} \cdot \frac{1}{2}\left(|V_2^{sp}|^2 - |V_{g2}|^2\right) - \frac{1}{2}\left(\sum_{m=1}^{n-1} V_2[m] V_2^*[n-m]\right) \end{bmatrix} \qquad \text{(A11)}$$

## REFERENCES

[1] B. Gao, G. K. Morison and P. Kundur, "Voltage stability evaluation using modal analysis," *IEEE Trans. Power Syst.*, vol. 7, no. 4, pp. 1529-1542, Nov. 1992.

[2] Y. Mansour, W. Xu *et al.*, "SVC placement using critical modes of voltage instability," *IEEE Trans. Power Syst.*, vol. 9, no. 2, pp. 757-763, May 1994.

[3] T. Van Cutsem and C. Vournas, *Voltage Stability of Electric Power Systems*, Norwell, MA: Kluwer, 1998.

[4] P. Löf *et al.*, "Fast calculation of a voltage stability index," *IEEE Trans. Power Syst.*, vol. 7, no. 1, pp. 54-64, Feb. 1992.

[5] J. M. Lim and C. L. DeMarco, "SVD-based voltage stability assessment from phasor measurement unit data," *IEEE Trans. Power Syst.*, vol. 31, no. 4, pp. 2557-2565, Jul. 2016.

[6] S. Greene *et al.*, "Sensitivity of the loading margin to voltage collapse with respect to arbitrary parameters," *IEEE Trans. Power Syst.*, vol. 12, no. 1, pp. 262-272, Feb. 1997.

[7] F. Capitanescu *et al.*, "Unified sensitivity analysis of unstable or low voltages caused by load increases or contingencies," *IEEE Trans. Power Syst.*, vol. 20, no. 1, pp. 321-329, Feb. 2005.

[8] T. Gou and R. Schlueter, "Identification of generic bifurcation and stability problems in power system differential-algebraic model," *IEEE Trans. Power Syst.*, vol. 9, pp. 1032-1044, May 1994.

[9] W. Marszalek and Z. Trzaska, "Singularity-induced bifurcations in electrical power systems," *IEEE Trans. Power Syst.*, vol. 20, pp. 312-320, Feb. 2005.

[10] V. Ajjarapu and C. Christy, "The continuation power flow: a tool for steady state voltage stability analysis," *IEEE Trans. Power Syst.*, vol. 7, no. 1, pp. 416–423, Feb. 1992.

[11] B. Wang, C. Liu and K. Sun, "Multi-stage holomorphic embedding method for calculating the power-voltage curve" *IEEE Trans. Power Syst.*, accepted.

[12] K. Vu *et al.*, "Use of local measurements to estimate voltage-stability margin," *IEEE Trans. Power Syst.*, vol. 14, no. 3, pp. 1029-1035, Aug. 1999.

[13] B. Milosevic and M. Begovic, "Voltage-stability protection and control using a wide-area network of phasor measurements," *IEEE Trans. Power Syst.*, vol. 18, no. 1, pp. 121–127, Feb. 2003.

[14] M. Glavic and T. Van Cutsem, "Wide-area detection of voltage instability from synchronized phasor measurements. Part I: Principle," *IEEE Trans. Power Syst.*, vol. 24, no. 3, pp. 1408–1416, Aug. 2009.

[15] F. Hu, K. Sun *et al*. "Measurement-based real-time voltage stability monitoring for load areas", *IEEE Trans. Power Systems*, vol. 31, no. 4, pp. 2787-2798, Jul. 2016.

[16] A. Trias, "The Holomorphic Embedding Load Flow Method," *IEEE PES GM,* San Diego, CA, Jul. 2012.

[17] A. Trias, "System and method for monitoring and managing electrical power transmission and distribution networks," US Patents 7 519 506 and 7 979 239, 2009–2011.

[18] A. Trias, "Fundamentals of the holomorphic embedding load–flow method" *arXiv: 1509.02421v1*, Sep. 2015.

[19] S. S. Baghsorkhi and S. P. Suetin, "Embedding AC Power Flow with Voltage Control in the Complex Plane: The Case of Analytic Continuation via Padé Approximants," *arXiv:1504.03249*, Mar. 2015.

[20] M. K. Subramanian, Y. Feng and D. Tylavsky, "PV bus modeling in a holomorphically embedded power-flow formulation," *North American Power Symposium (NAPS)*, Manhattan, KS, 2013.

[21] M. K. Subramanian, "Application of Holomorphic Embedding to the Power-Flow Problem," *Master Thesis*, Arizona State Univ., Aug. 2014.

[22] A. Trias, "Fundamentals of the Holomorphic Embedding Load-Flow Method," *arXiv.org*: 1509.02421, Sep. 2015.

[23] I. Wallace *et al.*, "Alternative PV bus modelling with the holomorphic embedding load flow method," *arXiv:1607.00163*, Jul. 2016

[24] S. S. Baghsorkhi and S. P. Suetin, "Embedding AC power flow in the complex plane part I: modelling and mathematical foundation," *arXiv:1604.03425*, Jul. 2016

[25] S. Rao, Y. Feng, D. J. Tylavsky and M. K. Subramanian, "The holomorphic embedding method applied to the power-flow problem," *IEEE Trans. on Power Syst.*, vol. 31, no. 5, pp. 3816-3828, Sep. 2016.

[26] S. Rao, D. J. Tylavsky and Y. Feng, "Estimating the saddle-node bifurcation point of static power systems using the holomorphic embedding method," *International Journal of Electrical Power and Energy Systems*, vol. 84, no. x, pp. 1-12, 2017.

[27] A. Trias and J. L. Marin, "The holomorphic embedding loadflow method for DC power systems and nonlinear DC circuits," *IEEE Trans. on Circuits and Systems-I: regular papers*, vol. 63, no.2, pp. 322-333, Sep. 2016.

[28] Y. Feng and D. Tylavsky, "A novel method to converge to the unstable equilibrium point for a two-bus system," *North American Power Symposium (NAPS)*, Manhattan, KS, 2013.

[29] Y. Feng, "Solving for the low-voltage-angle power-flow solutions by using the holomorphic embedding method," *Ph.D. Dissertation*, Arizona State Univ., Jul. 2015.

[30] S. Rao and D. Tylavsky, "Nonlinear network reduction for distribution networks using the holomorphic embedding method," *North American Power Symposium (NAPS)*, Denver, CO, USA, 2016.

[31] S. S. Baghsorkhi and S. P. Suetin, "Embedding AC power flow in the complex plane part II: a reliable framework for voltage collapse analysis," *arXiv:1609.01211*, Sep. 2016

[32] B Schmidt, "Implementation and evaluation of the holomorphic embedding load flow method," *Master Thesis*, Technical Univ. Munchen, Mar. 2015.

[33] M. R. Range, *Holomorphic Functions and Integral Representations in Several Complex Variables*, Springer-Verlag, New York, Inc. 2002.

[34] J. Nocedal and S. J. Wright, *Numerical Optimization*, 2nd ed. New York, NY, USA: Springer, 2006.

[35] A. Abur and A. G. Exposito, *Power System State Estimation*, Marcel Dekker, New York, Inc. 2004.

[36] H. Stahl, "On the convergence of generalized Pade approximants," *Constructive Approximation*, vol. 5, pp. 221-240, 1989.

[37] H. Stahl, "The convergence of Pade approximants to functions with branch points," *Journal of Approximation Theory*, vol. 91, no. 2, pp. 139-204, 1997.

[38] A. Cuyt *et al. Handbook of Continued Fractions for Special Functions*, Springer, ISBN: 978-1-4020-6948-2.

[39] J. Grainger and W. Stevenson, *Power System Analysis*, pp. 337-338, McGraw-Hill, 1994.

[40] DIgSILENT/PowerFactory, *Technical Reference Documentation—External Grid*.

[41] R. D. Zimmerman *et al.*, "MATPOWER: Steady-state operations, planning, and analysis tools for power systems research and education," *IEEE Trans. Power Syst.*, vol. 26, no. 1, pp. 12–19, Feb. 2011.

[42] J. H. Chow *et al.*, "Inertial and slow coherency aggregation algorithms for power system dynamic model reduction," *IEEE Trans. Power Syst.*, vol. 10, no. 2, pp. 680-685, May 1995.